\documentclass[11pt]{article}

\usepackage{caption}
\usepackage{epsfig}
\usepackage{color}
\usepackage{psfrag}
\usepackage{verbatim}
\usepackage{amsopn}
\usepackage{appendix}
\usepackage{dsfont,amsmath,amssymb,color,units,pstricks}
\usepackage{amsfonts}
\usepackage{graphicx}
\usepackage{natbib}
\usepackage{array}
\usepackage{lscape}
\usepackage{fullpage}
\usepackage{tikz}
\usepackage{url}
\usepackage{subcaption}
\allowdisplaybreaks
\usetikzlibrary{matrix}

\newcommand\be{\begin{equation}}
\newcommand\bea{\begin{eqnarray} \nonumber }
\newcommand\ee{\end{equation}}
\newcommand\eea{\end{eqnarray}}

\begin{document}

\unitlength = 1mm
\title{Latency and Liquidity Provision in a Limit Order Book}

\author{Julius Bonart$^{1,2}$, Martin D. Gould$^{1}$}

\maketitle

\noindent\small{$1$: CFM--Imperial Institute of Quantitative Finance, Department of Mathematics, Imperial College, London SW7 2AZ; $2$: Department of Computer Science, University College London, London WC1E 6BT}\\ 

\begin{abstract}We use a recent, high-quality data set from Nasdaq to perform an empirical analysis of order flow in a limit order book (LOB) before and after the arrival of a market order. For each of the stocks that we study, we identify a sequence of distinct phases across which the net flow of orders differs considerably. We note some of our results are consist with the widely reported phenomenon of stimulated refill, but that others are not. We therefore propose alternative mechanical and strategic motivations for the behaviour that we observe. Based on our findings, we argue that strategic liquidity providers consider both adverse selection and expected waiting costs when deciding how to act.\end{abstract}

\section{Introduction}\label{sec:intro}

The widespread uptake of electronic trading has facilitated a dramatic change in the way that traders supply and demand liquidity. In most modern financial markets, trade occurs via a continuous double-auction mechanism called a limit order book (LOB) \citep{Gould13}, in which traders interact by submitting two types of orders: market orders, which consume liquidity, and limit orders, which supply it. All traders can choose freely between submitting market orders or limit orders, and therefore between the provision or consumption of liquidity.

Before the widespread adoption of LOBs, liquidity provision was typically performed by a small group of designated specialists. These specialists determined the prices at which they were willing to buy or sell an asset, then communicated these prices to other traders in the market. All other traders who wished to buy or sell could only do so by transacting with a specialist at their advertised price. Therefore, the small group of designated specialists served as the exclusive source of liquidity for the whole market. In an LOB, by contrast, liquidity provision is a self-organized process driven by aggregate order flow. In this way, the temporal evolution of an LOB can be regarded as a dynamic feedback loop between the provision of liquidity and the execution of trades.

During the past decade, many publications have addressed questions about how liquidity influences the arrivals of market orders, and have thereby highlighted an important conditioning known as \emph{selective liquidity taking}, by which traders carefully select the size of their market orders according to the liquidity available in the LOB. Similarly, several LOB models have illustrated how selective liquidity taking may account non-trivial market phenomena such as the unpredictable nature of price changes and the concavity of price impact (see \citet{Bouchaud:2009digest} for a recent survey).

Despite this large literature that addresses how liquidity influences market order arrivals, relatively few publications to date have addressed the other direction in the feedback loop --- namely, how market order arrivals influence liquidity. Understanding this process is important for the several reasons. First, liquidity is directly related to the impact costs experienced by traders when buying or selling an asset. Given the considerable effort that many practitioners dedicate to minimizing such costs (see, e.g., \citet{BertsimasLo,AlmgrenChriss}), obtaining a better understanding of these dynamics is a task of high practical relevance. Second, understanding market resilience (i.e., the speed with which markets revert to their previous state after the arrival of a large market order) requires detailed understanding of the corresponding dynamics of liquidity provision. Third, insufficient liquidity provision can cause markets to become unstable. In recent years, several high-profile events, often referred to as ``flash crashes'' \citep{menkveld,KirilenkoKyle} have arisen from short-term liquidity crises, and have disturbed the normal-functioning of financial markets. Therefore, understanding how market order arrivals impact liquidity provision is an important task for market regulators seeking to understand the sources of market instabilities.

In this paper, we perform an empirical study of how market order arrivals impact LOB liquidity for 5 large-tick stocks on Nasdaq. Specifically, we calculate the mean net flow of limit orders at the same-side and opposite-side best quotes after the arrival of a market order. In each case, we identify a sequence of distinct phases across which the net flow of orders differs considerably. We find that the progression from each phase to the subsequent phases happens approximately contemporaneously for each of the stocks that we study.

After the arrival of a market order, we first observe a period that we call the \emph{platform-latency phase,} during which net order flow is exactly 0. Platform latency occurs due to the time it takes to process and route messages inside an automated trading platform \citep{Kirilenko:2015latency}. At both the same-side and opposite-side best quotes, we then observe a period in which net order flow is positive but very small. We call this phase the \emph{response-latency phase}, because the limit orders received by the server during this period are likely to have been submitted by their owners without knowledge of the previous market order arrival, and therefore do not reflect the new market order's arrival. After the response-latency phase, we observe a period that we call the \emph{high-speed reaction phase}, during which the net order flow is strongly negative at the same-side best quote and strongly positive at the opposite-side best quote. We argue that the behaviour in the high-speed reaction phase is attributable to high-frequency traders reacting to the market order arrival. At the same-side best quote, the high-speed reaction phase is followed by a strong net inflow of new orders, which is consistent with the widely observed phenomenon of stimulated refill (see e.g., \citet{Bouchaud:2006random,Gerig:2007theory,Rosenau}), by which the arrival of a market order encourages other traders to submit new limit orders at the same price. At the opposite-side best quote, the high-speed reaction phase is followed by a strong net outflow of limit orders, which we note is consistent with the hypothesis that liquidity providers consider the expected waiting costs of remaining in a limit order queue.

We also perform a similar empirical analysis of net order flow immediately before the arrival of a market order, and we again identify a sequence of distinct phases with different net order-flow patterns. At the same-side best quote, we find that net order flow is negative from at least 10 seconds before the upcoming market order arrival. This negative net flow is caused by traders cancelling their limit orders at the best quote before the market order arrival, which we note is consistent with the hypothesis that these traders fear adverse selection. This outflow of orders continues until shortly before the market order arrives, at which point the trend reverses and the net flow becomes slightly positive. We argue that this small positive inflow of new limit orders immediately before the market order suggests that some traders who submit market orders implement selective liquidity taking strategies to minimize their market impact. At the opposite-side best quote, the pattern is inverted: the net flow of orders is initially positive, then suddenly becomes negative. We discuss this result in the context of several recent empirical studies that propose \emph{order-book imbalance} (see, e.g., \citet{Avellaneda:2011forecasting,Donnelly,Gould:2015imbalance}) to be a strong predictor of future order flow, and we argue that the rapid change in order flow just before the market order arrival is caused by traders who cancel a limit order at the opposite-side best quote to instead perform the same trade (which then occurs at the same-side best quote) by submitting a market order.

The complex inflow and outflow of limit orders that we observe suggests that traders consider both adverse selection and expected waiting costs when deciding how to act. We thereby conclude that a realistic theory of strategic liquidity provision should include multiple interacting mechanisms on different timescales to reproduce the complex dynamics that arise empirically in modern financial markets.

The paper proceeds as follows. In Section \ref{sec:lobs}, we provide a detailed introduction to the mechanics of trading via LOBs. In Section \ref{sec:liquidityprovision}, we discuss the dynamic interactions between liquidity provision and consumption in an LOB and review several models and explanations of strategic liquidity provision developed elsewhere in the literature. In Section \ref{sec:data}, we discuss our data. In Section \ref{sec:methodology}, we discuss our methodology. We present our empirical results in Section \ref{sec:results} and discuss our findings in Section \ref{sec:discussion}. Section \ref{sec:conclusion} concludes.

\section{Limit Order Books}\label{sec:lobs}

More than half of the world's financial markets utilize electronic limit order books (LOBs) to facilitate trade \citep{Rosu09}. In an LOB, institutions interact via the submission of orders. An \emph{order} $x$ with price $p_x$ and size $\omega_x>0$ (respectively, $\omega_x<0$) is a commitment by its owner to sell (respectively, buy) up to $\left|\omega_x\right|$ units of the asset at a price no less than (respectively, no greater than) $p_x$.

Whenever an institution submits a buy (respectively, sell) order $x$, an LOB's trade-matching algorithm checks whether it is possible for $x$ to \emph{match to} an active sell (respectively, buy) order $y$ such that $p_y \leq p_x$ (respectively, $p_y \geq p_x$). If so, the matching occurs immediately and the owners of the relevant orders agree a trade for the specified amount at the specified price. If not, then $x$ becomes \emph{active} at the price $p_x$, and it remains active until it either matches to an incoming sell (respectively, buy) order or is \emph{cancelled} by its owner. Orders that match upon arrival are called \emph{market orders}. Orders that do not match upon arrival are called \emph{limit orders,} and become \emph{active}. An LOB $\mathcal{L}(t)$ is the collection of all active orders for a given asset on a given platform at a given time $t$.

Institutions trading in an LOB can choose freely between submitting market orders or limit orders. Market orders are certain to match immediately, but never do so at a price better than $b_t$ or $a_t$. Conversely, limit orders may eventually match at better prices than market orders, but their execution is uncertain because it depends on the arrival of a future market order of opposite type. In short, limit orders stand a chance of matching at better prices than do market orders, but they also run the risk of never being matched.

Throughout this paper, we use the term \emph{liquidity provision} to describe the submission of limit orders, which create the possibility for future transactions in the market. Similarly, we use the term \emph{liquidity consumption} to describe the submission of market orders, which trigger transactions by executing previously submitted limit orders.

\begin{figure}
\centering
\includegraphics[width=0.5\textwidth]{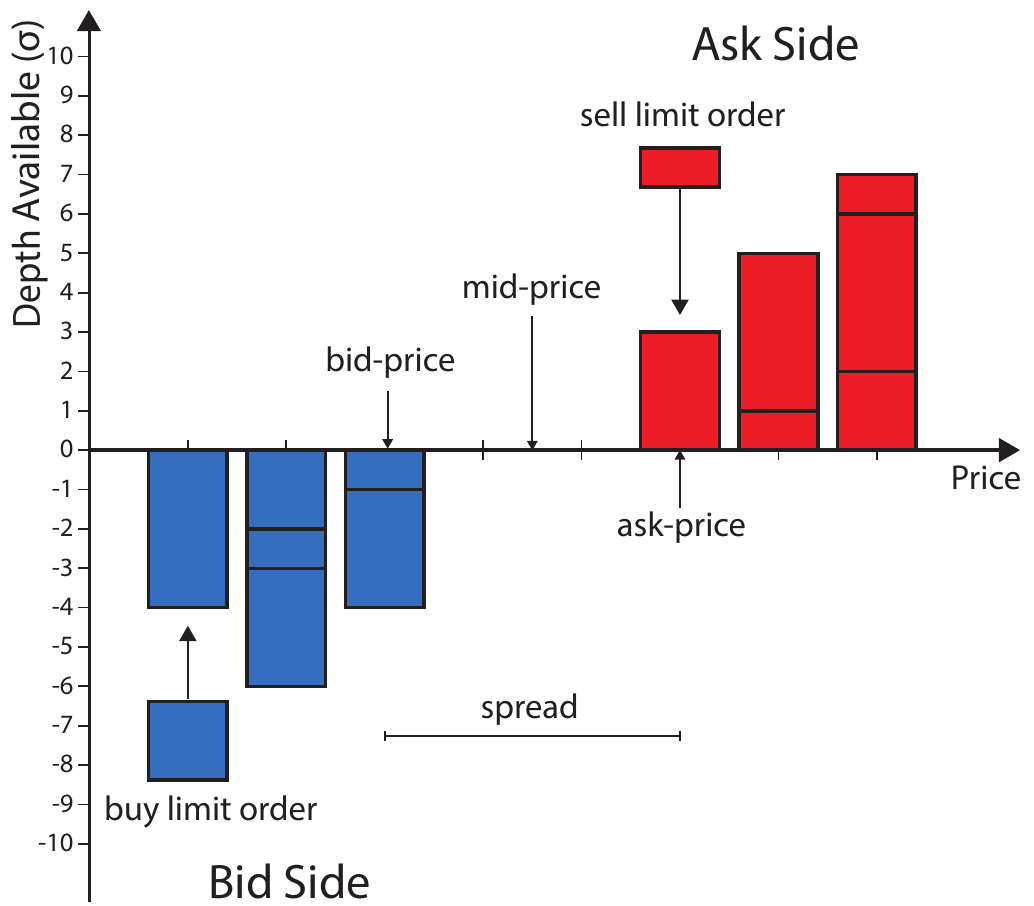}
\caption{Schematic of an LOB. The horizontal lines within the blocks at each price level denote the different active orders at each price.}
\label{fig:schematic}
\end{figure}

In an LOB, the \emph{bid price} $b_t$ is the highest price among active buy orders at time $t$. Similarly, the \emph{ask price} $a_t$ is the lowest price among active sell orders at time $t$. The bid and ask prices are collectively known as the \emph{best quotes}. Their difference $s_t=a_t-b_t$ is called the \emph{bid-ask spread,} and their mean $m_t=(a_t+b_t)/2$ is called the \emph{mid price}. For a given price $q$ and time $t$, we say that $q$ is \emph{on the buy side} if $q \leq b_t$, that $q$ is \emph{on the sell side} if $q \geq a_t$, or that $q$ is \emph{inside the spread} if $b_t < q < a_t$. Figure \ref{fig:schematic} shows a schematic of an LOB at some instant in time, illustrating these definitions.

When trading in an LOB, institutions must choose the price of their orders according to the platform's \emph{tick size} $\pi>0$, which is the smallest permissible price interval between different orders. All orders must arrive with a price that is an integer multiple of the tick size. Therefore, it is common for several different active orders to reside at the same price at a given time. To help traders evaluate the state of the market, electronic trading platforms typically summarize the information in $\mathcal{L}(t)$ by disseminating a feed that lists the aggregate quantities offered for purchase or sale at a set of price levels.

To determine the queueing priority among orders at a given price, most exchanges implement a \emph{price--time} priority rule. That is, for active buy (respectively, sell) orders, priority is given to the active orders with the highest (respectively, lowest) price, and ties are broken by selecting the active order with the earliest submission time.

The rules that govern order matching in an LOB also dictate how prices evolve through time. Consider a buy (respectively, sell) order $x$ that arrives immediately after time $t$. If $p_x \leq b_t$ (respectively, $p_x \geq a_t$), then $x$ is a limit order that becomes active upon arrival and does not cause $b_t$ or $a_t$ to change. If $b_t < p_x < a_t$, then $x$ is a limit order that becomes active upon arrival and causes $b_t$ to increase (respectively, $a_t$ to decrease) to $p_x$. If $p_x \geq a_t$ (respectively, $p_x \leq b_t$), then $x$ is a market order that matches to one or more active sell (respectively, buy) orders upon arrival. When such a matching occurs, it does so at the price of the active order, which is not necessarily equal to $p_x$. Whether or not such a matching causes $a_t$ (respectively, $b_t$) to change depends on whether or not $\left|\omega_x\right|$ exceeds the total size available for sale at $a_t$ (respectively, for purchase at $b_t$). Price changes also occur if the total volume available for sale at $a_t$ (respectively, for purchase at $b_t$) is cancelled.

\section{Liquidity Provision in a Limit Order Book}\label{sec:liquidityprovision}

In an LOB, the queueing dynamics associated with the arrivals and departures of orders emerge from the provision and consumption of liquidity. Several authors from a wide range of disciplines have sought to explain LOB queue dynamics in this context. Such investigations have taken a variety of different starting points, drawing on ideas from economics, physics, mathematics, statistics, and psychology. In this section, we review a selection of publications most relevant to our work. For detailed surveys, see \citet{Bouchaud:2009digest}, \citet{Gould13}, or \citet{Chakraborti:2011empirical,Chakraborti:2011agent}.

The earliest models of LOBs typically described order flow according to simple stochastic processes with fixed rate parameters. \citet{Smith:2003statistical} introduced a model in which limit order arrivals, market order arrivals, and cancellations occur as mutually independent Poisson processes with fixed rate parameters. \citet{ContStoikov} extended this model by allowing the rates of limit order arrivals and cancellations to vary across prices. \citet{Mike:2008empirical} incorporated long-range autocorrelation effects between the signs (buy/sell) of successive orders. Subsequently, \citet{ContLarrard} and \citet{Gareche13} also studied similar LOB models, but only considered the dynamics of the queues at the best quotes.

Although these so-called ``zero-intelligence'' models\footnote{The term ``zero intelligence'' is used to describe models in which aggregated order flows are assumed to be governed by specified stochastic processes. In this way, order flow can be regarded as a consequence of traders blindly following a set of rules without strategic considerations.} of LOBs perform reasonably well at predicting some long-run statistical properties of real LOBs (see, e.g., \citet{Farmer:2005predictive}), their exclusion of explicit strategic considerations hinders their ability to make useful predictions about how liquidity providers might adapt their order flow according to the actions of other traders. Therefore, \citet{queue-reactive} recently sought to extend these queueing models by including strategic considerations, and thereby improved their predictive power.

When choosing how to act, institutions must weigh up the pros and cons of limit versus market order submissions. On the one hand, the possible price improvement offered by a limit order generates a clear incentive for institutions to provide liquidity. Indeed, some institutions submit buy and sell limit orders simultaneously, with the aim of earning their price difference if both orders are matched.\footnote{Many authors use the term ``market maker'' to describe an institution that performs this role. However, in the context of an LOB, this term does not imply that a given institution is a designated ``specialist'' with elevated status in the marketplace, as was the case for market makers in older, quote-driven markets.} On the other hand, liquidity providers expose themselves to risks that can severely harm their ability to earn profits. As we describe in the next section, several authors have highlighted how these considerations have caused liquidity provision in modern financial markets to become a highly sophisticated task in which liquidity providers dynamically adjust the amount of unmatched liquidity available.

In the existing literature on LOBs, there are two main theories for why liquidity providers' actions should depend on the behaviour of the other traders in an LOB. The first is that of \emph{information asymmetry}. In an early work on the topic, \citet{glosten} argued that submitting limit orders exposes institutions to the risk of \emph{adverse selection} from other institutions that have superior private information about the likely future price of the asset, and who thereby submit market orders to ``pick off'' mis-priced limit orders from less-well-informed institutions. \citet{glosten} argued that uninformed liquidity providers would factor in these adverse selection costs when choosing the prices for their limit orders, ultimately leading to a wider bid--ask spread. \citet{Chakravarty93} extended the \citet{glosten} model by acknowledging that informed traders could also implement complex strategies that involve submitting limit orders, not just market orders.

The second theory is that of \emph{execution uncertainty}, which arises because of the uncertain waiting time between the submission and execution of a limit order. \citet{Foucault05} and \citet{Rosu09} argued that execution uncertainty is an important determinant of LOB dynamics. \citet{Rosu14} noted that institutions who attempt to exploit private information about the likely future value of an asset experience an ``information slippage'' cost, because private information naturally becomes stale over time. \citet{Ohara86} and \citet{HO81} argued that inventory risk can also create waiting costs if a net position cannot be cleared sufficiently quickly. In models that consider execution uncertainty, the level of trader impatience often plays a central role in determining market dynamics. If the expected waiting time between the submission and execution of a limit order is large, then institutions tend to prefer the immediate execution associated with market orders. Conversely, if this expected waiting time is small, traders are more likely to tolerate the delayed execution in exchange for the opportunity to trade at a better price.

\section{Data}\label{sec:data}

Our empirical calculations are based on a data set that describes the LOB dynamics for 5 highly liquid stocks traded on Nasdaq during the six-month period of 1 March 2015 to 1 September 2015.\footnote{To ensure that our results are robust to the choice of time period, we also repeated our calculations using data from 1 March 2013 to 1 September 2013. We found that our results for this period were qualitatively similar to those for 1 March 2015 to 1 September 2015.} The data that we study originates from the LOBSTER database \citep{Huang:2011LOBSTER}, which lists all market order arrivals, limit order arrivals, and cancellations that occur on the Nasdaq platform during $09$:$30$ -- $16$:$00$ on each trading day. Trading does not occur on weekends or public holidays, so we exclude these days from our analysis. We also exclude market activity during the first and last $1000$ seconds of each trading day, to remove any abnormal trading behaviour that can occur shortly after the opening auction or shortly before the closing auction.

On the Nasdaq platform, each stock is traded in a separate LOB with price--time priority, with a tick size of $\pi = \$0.01$ (see Section \ref{sec:lobs}). Although this tick size is the same for all stocks on the platform, the prices of different stocks vary across several orders of magnitude (from about $\$1$ to more than $\$1000$). Therefore, the \emph{relative tick size} (i.e., the ratio between the stock price and $\pi$) similarly varies considerably across different stocks. In this paper, we restrict our attention to \emph{large-tick stocks}, for which the ratio between $\pi$ and the stock price is large. An important reason for doing so is that for large-tick stocks, the spread is very often equal to its minimum size $s_t=\pi$. When this occurs, one mechanism leading to a change in $b_t$ or $a_t$ is eliminated. Specifically, when $s_t=\pi$, institutions cannot submit limit orders inside the spread. Therefore, the only way in which $b_t$ or $a_t$ can change is if the order queue at either $b_t$ or $a_t$ depletes to zero. For small-tick stocks, by contrast, $s_t$ is usually larger than $\pi$, so any institution can submit a buy (respectively, sell) limit order inside the spread, and thereby cause $b_t$ to increase (respectively, $a_t$ to decrease). The arrival of many limit orders inside the spread could obscure the queue dynamics that we seek to investigate.

To choose the stocks in our sample, we first create a list of all stocks whose mid price remained below $\$50.00$ during the sample period of 1 March 2015 to 1 September 2015. We then order these stocks according to their total dollar trade value during this period, and select the first 5 stocks on this list. In descending order of their levels of market activity, these stocks are Microsoft (MSFT), Intel (INTC), Yahoo (YHOO), Micron Technology (MU), and Cisco (CSCO). Table~\ref{tab1} lists summary statistics describing trading activity for these 5 stocks during our sample period. 

\begin{table}
\begin{center}
\begin{small}
\begin{tabular}{|l|c|c|c|c|c|}
\hline
 & MSFT & INTC & YHOO & MU & CSCO \\
\hline
Total number of events at the best quotes & 59966351 & 31821544 & 29325492 & 25321792 & 24364686 \\
Percentage of market order arrivals & $1.7\%$ & $2.2\%$ & $2.4\%$ & $3.3\%$ & $1.9\%$ \\
Percentage of limit order arrivals & $52.9\%$ & $53.1\%$ & $52.7\%$ & $52.9\%$ & $51.7\%$ \\
Percentage of limit order cancellations & $45.4\%$ & $44.7\%$ & $44.9\%$ & $43.8\%$ & $46.4\%$ \\
Mean bid--ask spread [$\$$] & $0.0117$ & $0.0118$ & $0.0122$ & $0.0127$ & $0.0115$ \\
Mean trade price [$\$$] & $45.14$ & $31.27$ & $41.60$ & $24.49$ & $28.42$ \\
Mean volume at the best quotes [shares] & 5131 & 6740 & 2092 & 3514 & 11423\\
Mean size of market orders [shares]& 617 & 742 & 361 & 569 & 857\\
Mean size of price-maintaining market orders & 455 & 520 & 269 & 383 & 625\\
   \hline
\end{tabular}
\caption{Summary statistics for the 5 stocks in our sample.}
\label{tab1}
\end{small}
\end{center}
\end{table}

The LOBSTER data has many important benefits that make it particularly suitable for our study. First, the data is recorded directly at the Nasdaq servers. Therefore, we avoid the many difficulties associated with data sets that are recorded by third-party providers, such as misaligned or inaccurate time stamps and incorrectly ordered events. Second, each market order arrival listed in the data contains an explicit identifier for the limit order to which it matches. This enables us to perform one-to-one matching between market and limit orders, without the need to apply inference algorithms for this purpose, which can produce noisy and inaccurate results. Third, each limit order described in the data constitutes a firm commitment to trade. Therefore, our results reflect the market dynamics for real trading opportunities, not ``indicative'' declarations of intent. Fourth, each LOB event is recorded with a time stamp in nanoseconds. This enables us to consider market activity that occurs very soon after the arrival of a market order and provides an extremely detailed level of granularity when tracking the temporal evolution of net order flow.

The LOBSTER database describes all LOB activity that occurs on Nasdaq, but does not provide any information regarding order flow for the same assets on different platforms. To minimize the possible impact on our results, we restrict our attention to stocks for which Nasdaq is the primary trading venue. Despite the advanced fragmentation of today's equity markets, Nasdaq captures $42\%$ of the total trading volume for MSFT, $45\%$ for INTC, $41\%$ for CSCO, $31\%$ for YHOO, and $33\%$ for MU. Our results enable us to identify several robust statistical regularities in the net order flow before and after market order arrivals, which is precisely the aim of our study. We therefore do not regard this feature of the LOBSTER data to be a serious limitation for the present study.

\section{Methodology}\label{sec:methodology}

The aim of our empirical calculations is to quantify how liquidity providers react to the arrival of market orders. To do so, we use the LOBSTER data (see Section \ref{sec:data}) to calculate the temporal evolution of the bid and ask volumes around each such event.

Let the \emph{bid volume} $V^B(t)$ denote the total size of active buy orders at the bid price $b_t$ at time $t$. Similarly, let the \emph{ask volume} $V^A(t)$ denote the total size of active sell orders at the ask price $a_t$ at time $t$. Throughout this paper, we use the time series $V^B(t)$ and $V^A(t)$ to study the cumulative net order flows at the best quotes. In this section, we describe our methodology for performing these calculations in situations when the values of the best quotes $b_t$ and $a_t$ do not change. This methodology forms the basis for our main empirical calculations throughout the paper. At the end of Section \ref{sec:results}, we relax this constraint to include cases in which the quote prices changes; we provide a detailed discussion of the corresponding methodology in Section \ref{subsec:dynamic_price}.

For a given stock on a given trading day, let $t_1, t_2, \ldots, t_N$ denote the times of the market order arrivals. For each $i=1,2,\ldots,N-1$ and a given $T \in \mathbb{R}$, we say that the $i^{\text{th}}$ market order arrival is \emph{$T$-separated} if $t_{i+1}-t_{t} \geq T$. We say that a buy (respectively, sell) market order is \emph{price maintaining} if its arrival does not consume the whole order queue at $a_t$ (respectively, $b_t$). Otherwise, we say that it is \emph{price changing.} An arriving buy (respectively, sell) market order is price maintaining if and only if it does not cause either $b_t$ or $a_t$ to change.

For a given stock and a given separation time $T>0$, let $\mathfrak M(T)$ denote the set of $T$-separated, price-maintaining market orders. We restrict our attention to the activity that occurs around the arrival of the market orders in $\mathfrak{M}(T)$. By considering only these market orders, we are able to concentrate on the dynamics of the bid and ask volumes without incorporating the effects of an initial price change and without considering market orders that arrive in rapid succession, both of which could make our results more difficult to interpret.

For a given market order arrival time $t_i$ and any $\tau \in \mathbb{R}^{>0}$, the \emph{buy-side cumulative net order flow} $W^B(t_i,\tau)$ measures the cumulative difference between the volumes of arriving and cancelled buy limit orders at $b_t$ during the time interval $\left(t_i,t_i+\tau\right]$,\begin{equation}\label{eq:WB}W^B(t_i,\tau) = V^B(t_i+\tau)-V^B(t_i).\end{equation}Similarly, the \emph{sell-side cumulative net order flow} $W^A(t_i,\tau)$ measures the cumulative difference between the volumes of arriving and cancelled sell limit orders at $a_t$ during the time interval $\left(t_i,t_i+\tau\right]$,\begin{equation}\label{eq:WA}W^A(t_i,\tau) = V^A(t_i+\tau)-V^A(t_i).\end{equation}Observe that $W^B$ and $W^A$ measure the cumulative net \emph{flow} of orders (i.e., the cumulative net arrivals of limit orders), not the absolute queue lengths.

\section{Results}\label{sec:results}

We now present our main empirical results. In Section \ref{subsec:separation}, we investigate the distribution of inter-arrival times for market orders. In Section \ref{subsec:after_same}, we investigate the temporal evolution of the mean net order flow at the same-side best quote. In Section \ref{subsec:after_opposite}, we repeat the same analysis at the opposite-side best quote. In Section \ref{subsec:before}, we investigate the temporal evolution of the mean net order flow at the same-side and opposite-side best quotes, but \emph{before} the arrival of a market order. In Section \ref{subsec:dynamic_price}, we discuss how our results from the other sections change if we relax the constraint that the values of the quotes are constant during the period of study.

\subsection{Inter-Arrival Times}\label{subsec:separation}

\begin{figure}
\centering
\includegraphics[width = 0.45\linewidth]{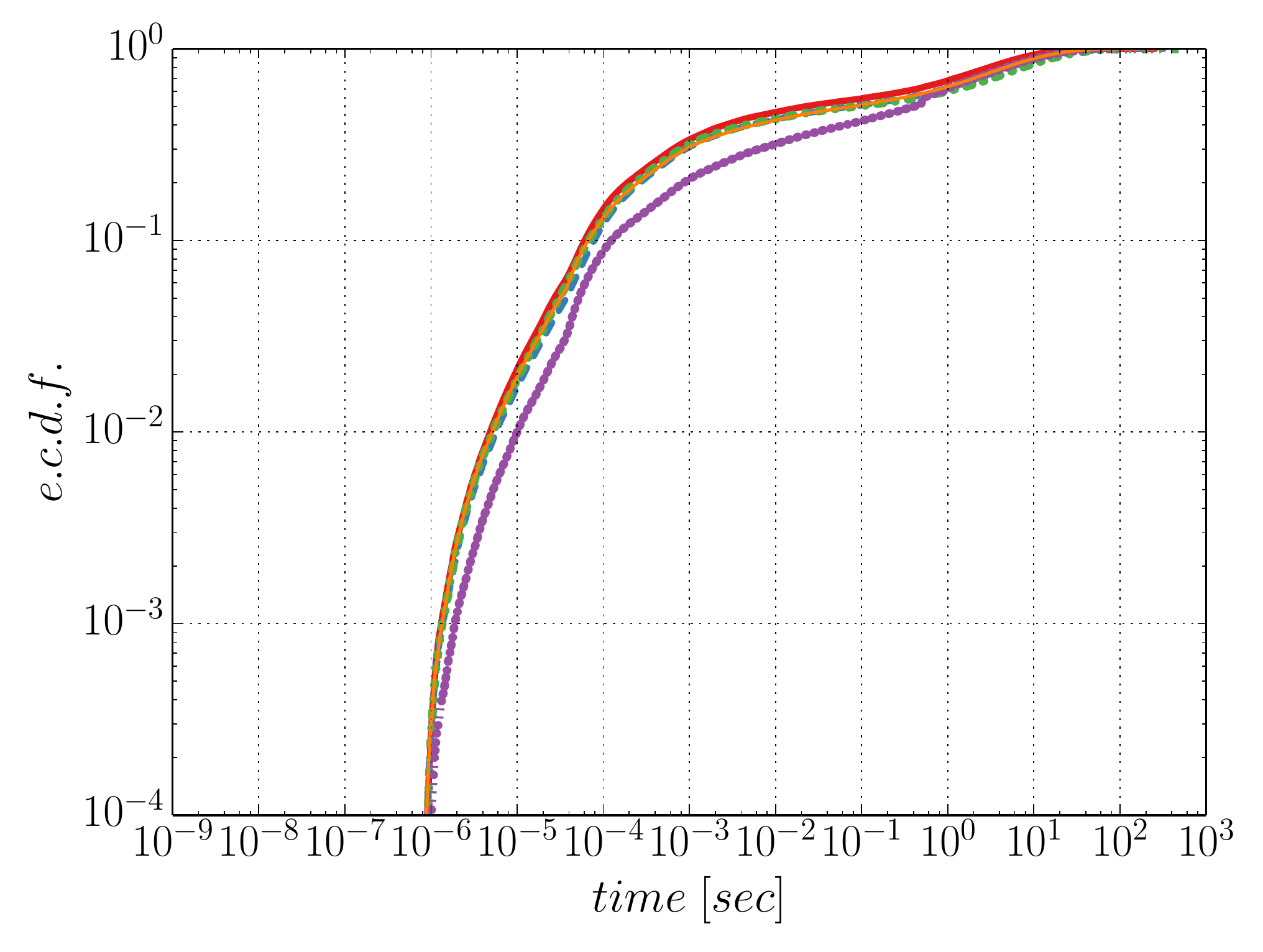}
\includegraphics[width = 0.45\linewidth]{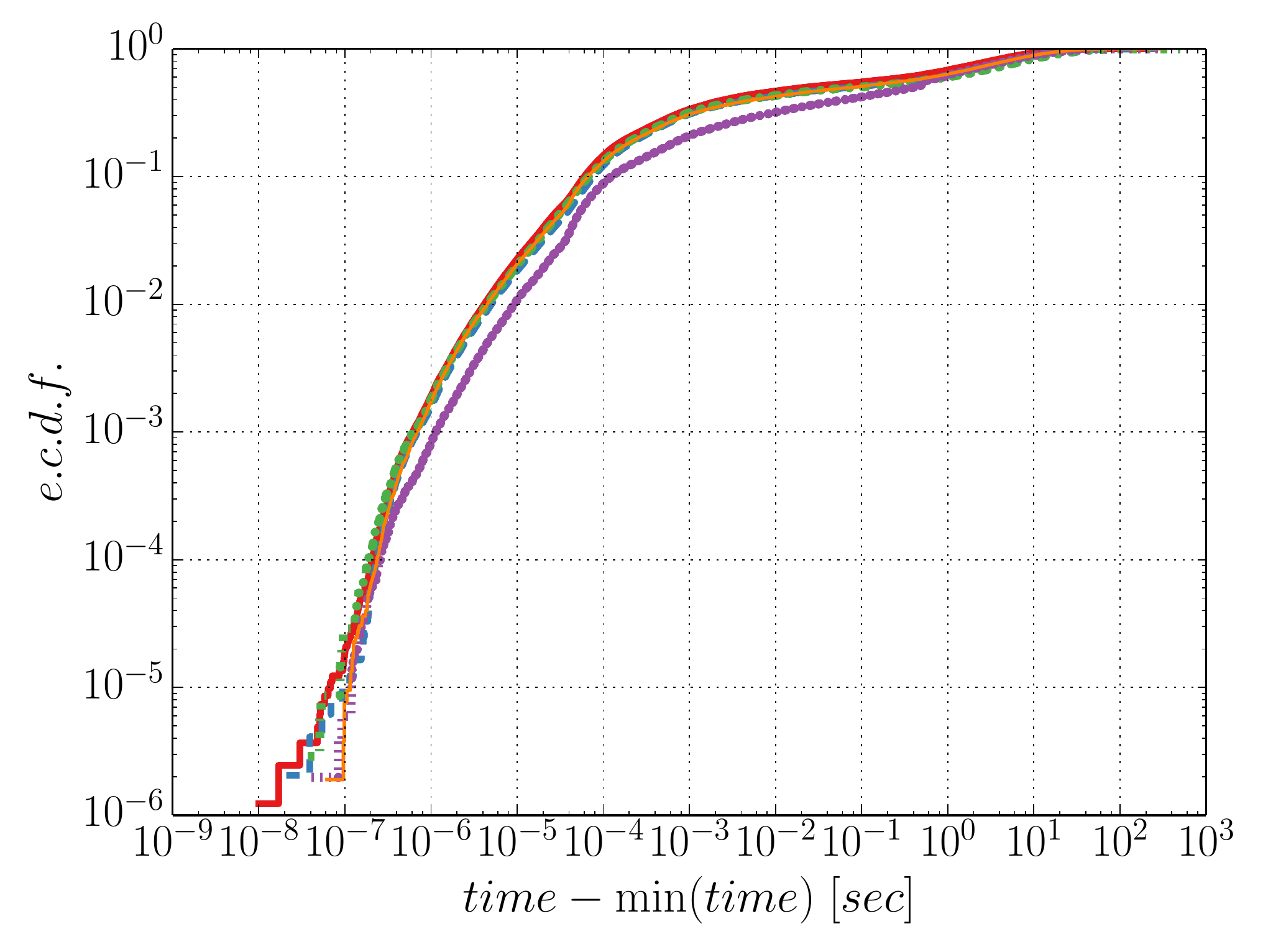}
\caption{Empirical cumulative density functions (ECDFs) of inter-arrival times $\Delta t_i = t_{i+1}- t_i$ for (solid red) MSFT, (dashed blue) INTC, (dash-dotted green) CSCO, (dotted violet) YHOO, and (thin solid orange) MU market order arrivals. The left panel shows the ECDF for the $\Delta t_i$ directly; the right panel shows the same plots after subtracting the minimum inter-arrival time for each stock (which is about $0.7 \times 10^{-6}$ seconds in each case).}
\label{fig:Deltat_lower}
\end{figure}

We first study the distribution of inter-arrival times of market orders in our sample. In the left panel of Figure \ref{fig:Deltat_lower}, we show the lower (i.e., short-time) tails of the empirical cumulative density functions (ECDF) of inter-arrival times $\Delta t_i = t_{i+1}- t_i$. Despite considerable differences in their aggregate trading activity (see Table \ref{tab1}), the shortest inter-arrival time that we observe is about $0.7 \times 10^{-6}$ seconds. This lower bound on inter-arrival times suggests that the shortest inter-arrival times do not reflect the differences in trading activity for the 5 stocks, but rather reflect the trading platform's internal latency, which occurs due to the time it takes to process and route trading messages. For a full discussion and empirical analysis of platform latency, see \citet{Kirilenko:2015latency}. 

The right panel of Figure \ref{fig:Deltat_lower} shows the same ECDFs after subtracting from each inter-arrival time the minimum value of $\Delta t_i$ for each stock. Each stock's ECDF has a qualitatively similar shape. About $10\%$ of market orders have an inter-arrival time of less than about $10^{-4}$ seconds, and the empirical distribution appears to scale approximately as a power law in this lower-tail (i.e., short-time) region.

\begin{figure}
\centering
\includegraphics[width = 0.45\linewidth]{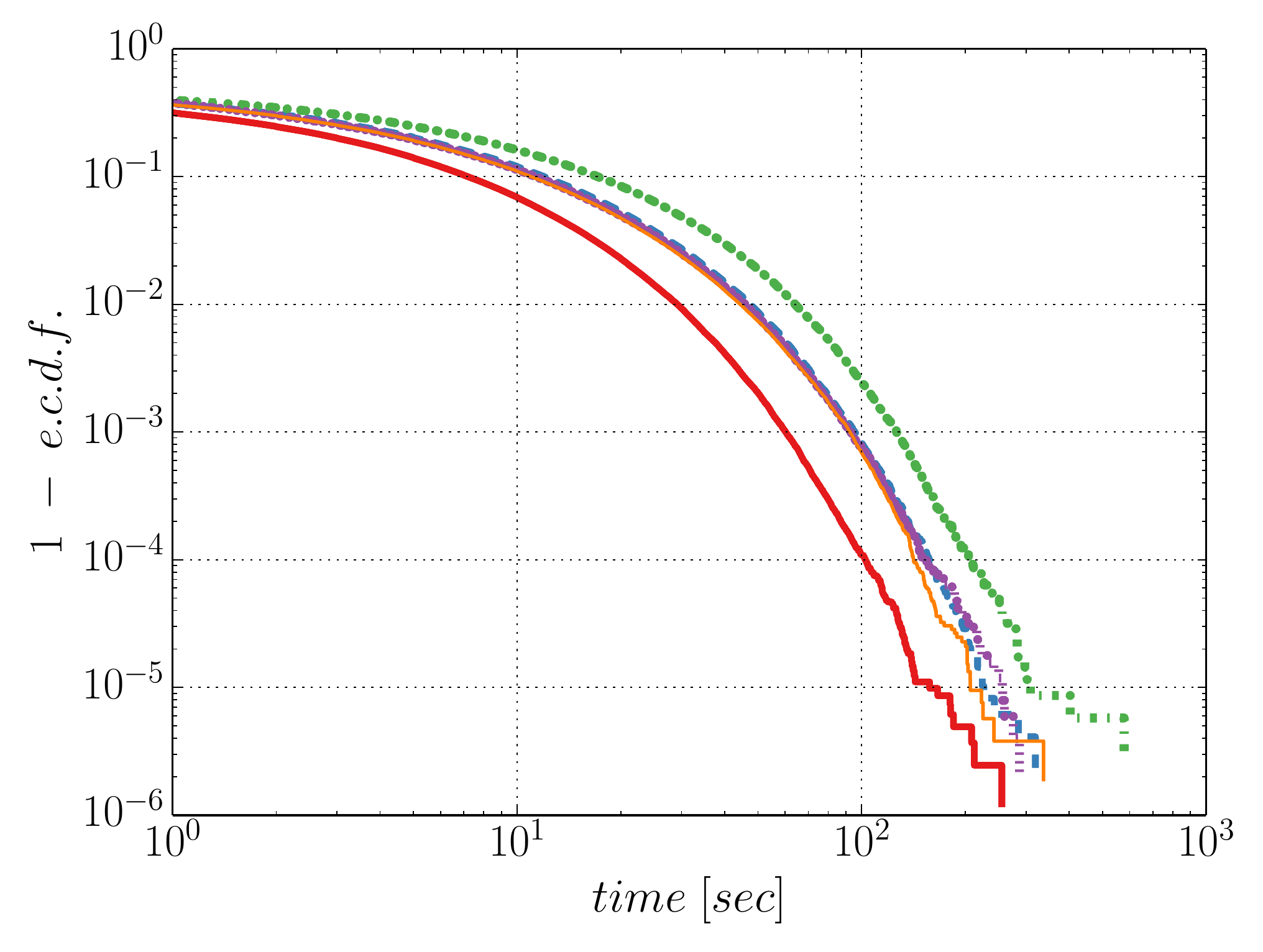}
\caption{Upper tails of the ECDFs of inter-arrival times $\Delta t_i = t_{i+1}- t_i$ for (solid red) MSFT, (dashed blue) INTC, (dash-dotted green) CSCO, (dotted violet) YHOO, and (thin solid orange) MU market order arrivals. Each plot shows 1 minus the ECDF in doubly logarithmic coordinates.}
\label{fig:Deltat_upper}
\end{figure}

To help illustrate the behaviour in the upper-tail (i.e., long-time) region, Figure \ref{fig:Deltat_upper} shows 1 minus the ECDF for each stock, in doubly logarithmic coordinates. The shape of the distribution is again similar for each stock, but the upper tails vary quantitatively across the different stocks. The most heavily traded stock (MSFT) lies below all other curves and the least heavily traded stock (CSCO) lies above all other curves. Together with Figure \ref{fig:Deltat_lower}, this suggests that the high-frequency inter-arrival times are quite similar across all stocks but that the lower-frequency inter-arrival times reflect the different levels of trading activity between the different stocks in our sample.

\subsection{Net Order Flow at the Same-Side Best Quote}\label{subsec:after_same}

\begin{figure}
  \centering
  \includegraphics[width = 0.45\linewidth]{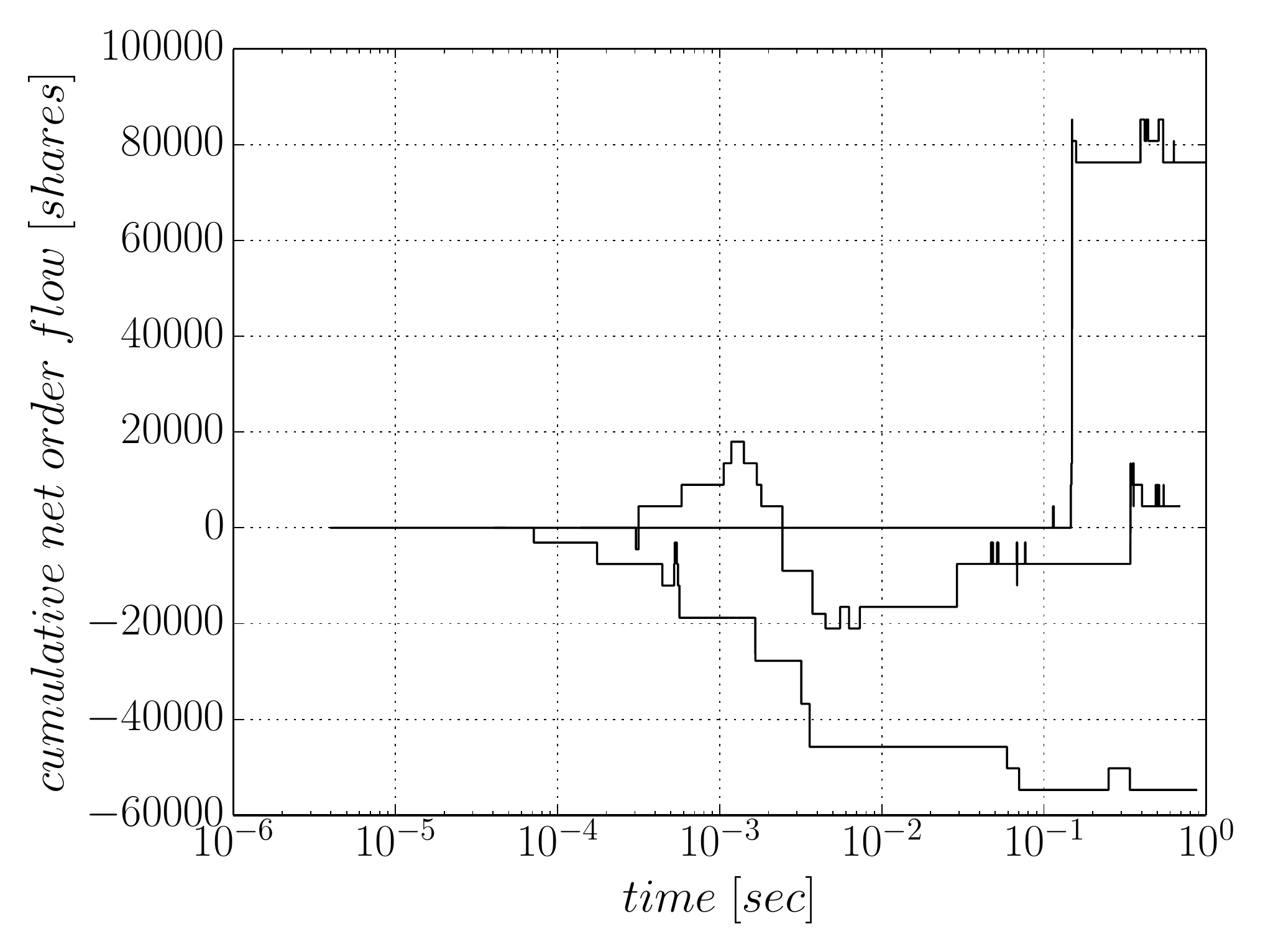}
  \includegraphics[width = 0.45\linewidth]{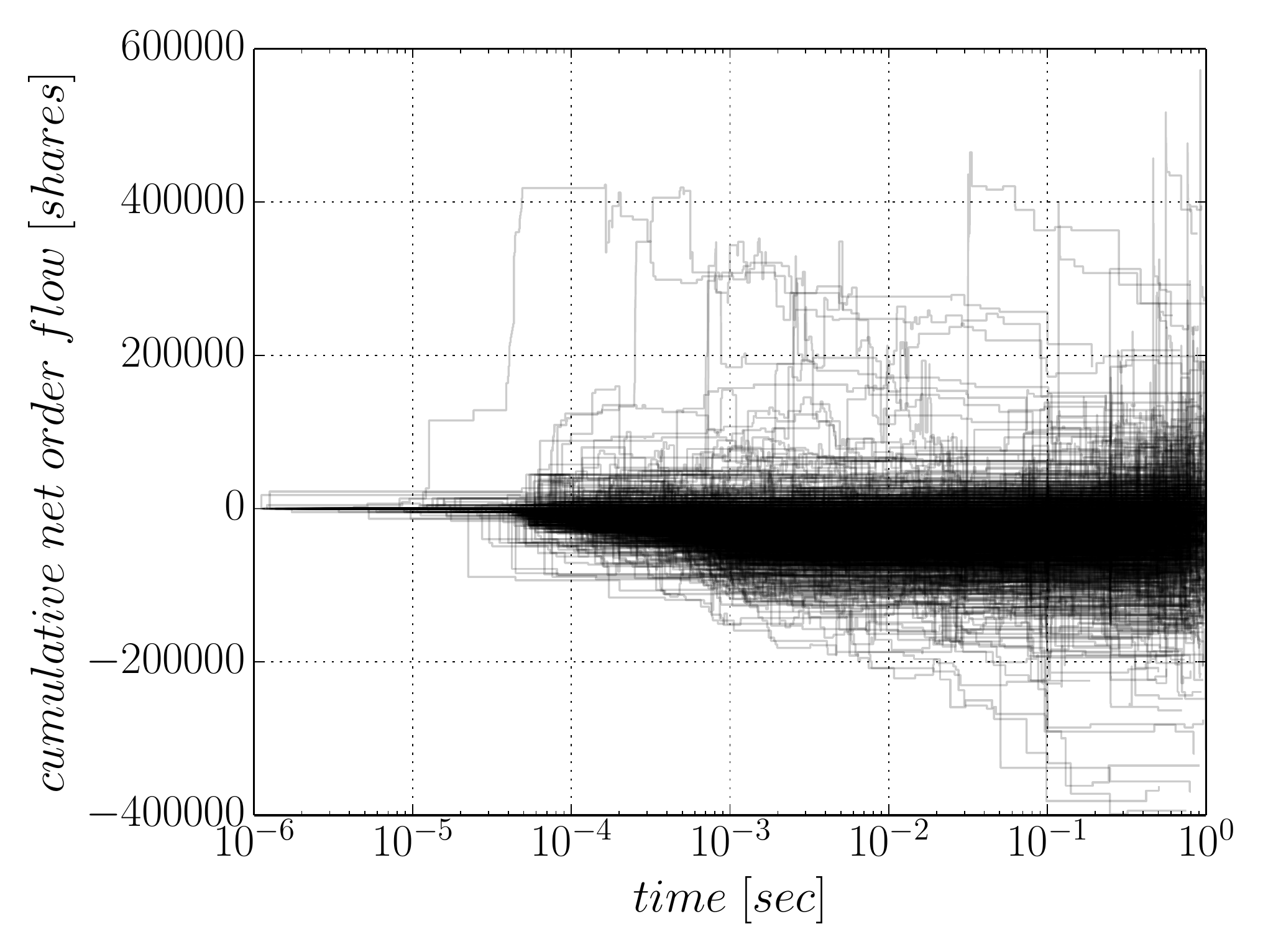}
  \caption{A selection of individual cumulative net order-flow trajectories at the same-side best quote for MSFT, after the arrival of a price-maintaining market order with a separation time of at least $T=1$ second. The left plot shows 3 trajectories that we chose uniformly at random; the right plot shows all the trajectories on 31 March 2015.}
  \label{fig:traj}
\end{figure}

Figure~\ref{fig:traj} shows a selection of cumulative net order-flow trajectories for MSFT, at the same-side best quote and after the arrival of a market order. Several features of these plots reveal interesting dynamics about the underlying order flow. First, individual trajectories can be extremely noisy and often contain short bursts of high-activity. Second, these short bursts are sometimes separated by long stretches of inactivity. Third, individual trajectories often contain rapid oscillations between a net inflow and a net outflow of limit orders.

Due to these three empirical properties of the order flow that we observe, it is difficult to develop a clear understanding of net order flow based on single trajectories alone. To help identify robust trends among these noisy observations, we therefore calculate the mean cumulative net order flow by averaging our results across the many different trajectories for each stock.

Specifically, for a given stock and a given time $\tau$, we calculate the mean cumulative net order flow $\tau$ seconds after the market order arrival, averaged across all $\tau$-separated market orders. In this way, if the arrival of market order $i+1$ occurs $2$ seconds after the arrival of market order $i$, then we include the $i^{\text{th}}$ cumulative net order flow trajectory in our averages for all time up to and including $2$ seconds, but not for times $\tau > 2$ seconds.

When calculating these averages, we align buy-side and sell-side activity by conditioning on the direction (i.e., buy/sell) of the arriving market order. In this way, we arrive at the \emph{same-side mean cumulative net order flow}\begin{equation}\label{eq:v_bar_same_after}\bar V^{s}(\tau) := \frac{1}{|\mathfrak M(\tau)|}\sum_{k \in \mathfrak M(\tau)} \left[ \mathbf 1_{\omega_k>0}  W^{B}(t_k,\tau) + \mathbf 1_{\omega_k<0} W^A(t_k,\tau)\right],\end{equation}where $\mathbf{1}$ denotes the indicator function.

We use a standard non-parametric bootstrap to estimate the standard error of our estimates at each $\tau$. Specifically, for each $l=1,2,\ldots,10000$, we draw a random sample (with replacement) of size $\left|\mathfrak M(\tau)\right|$ from the values of $W^{B}(t_k,\tau)$ and $W^{A}(t_k,\tau)$. We calculate the corresponding value of $\bar V^{s}(\tau)$ among this random sample, and we label this estimate $\bar V_l^{s}(\tau)$. We repeat this process for each $l$, using a different seed for the pseudo-random number generator in each case. Our estimate of the standard error of $\bar V^{s}(\tau)$ is the sample standard deviation of the $\bar V_l^{s}(\tau)$. We estimate the standard error of our other mean cumulative net order-flow trajectories throughout the paper in the same way.

As illustrated in Table~\ref{tab1}, the statistical properties of order flow vary considerably across the stocks in our sample. Therefore, in order to facilitate cross-stock comparisons, we also normalize our results for each stock. Specifically, we divide each stock's value of $\bar V^{s}(\tau)$ by the mean number of shares at the best quote for that stock, when averaged across our entire sample. There are also many other possible choices for this cross-stock normalization (such as dividing by the mean size of market orders or dividing by the mean number of shares at the best quote immediately before or after the arrival of a market order); we choose to normalize by the mean number of shares at the best quote for four reasons. First, this choice of normalization is intuitively appealing because it accounts for a given stock's liquidity: all else being equal, a more liquid stock will have a larger mean queue length. Second, the mean queue length varies considerably across the stocks in our sample and normalizing by this quantity helps to reduce the cross-stock variation in our results. Third, the mean queue length is easy to measure and simple to interpret. Fourth, the unit of mean queue length is ``shares'', so normalizing cumulative net order flow in this way produces a dimensionless quantity.

\begin{figure}
  \centering
	\includegraphics[width = 0.45\linewidth]{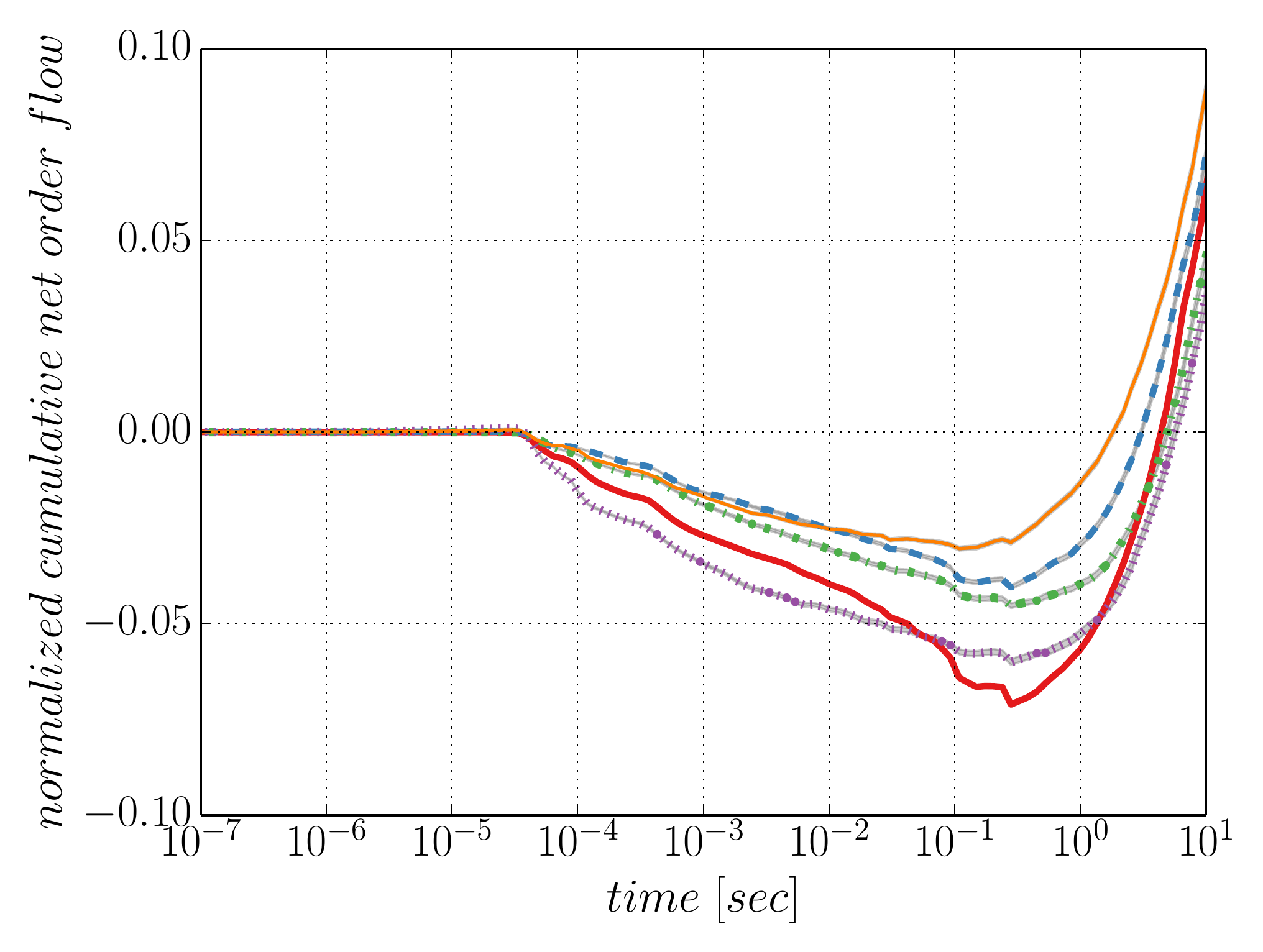}
  \includegraphics[width = 0.45\linewidth]{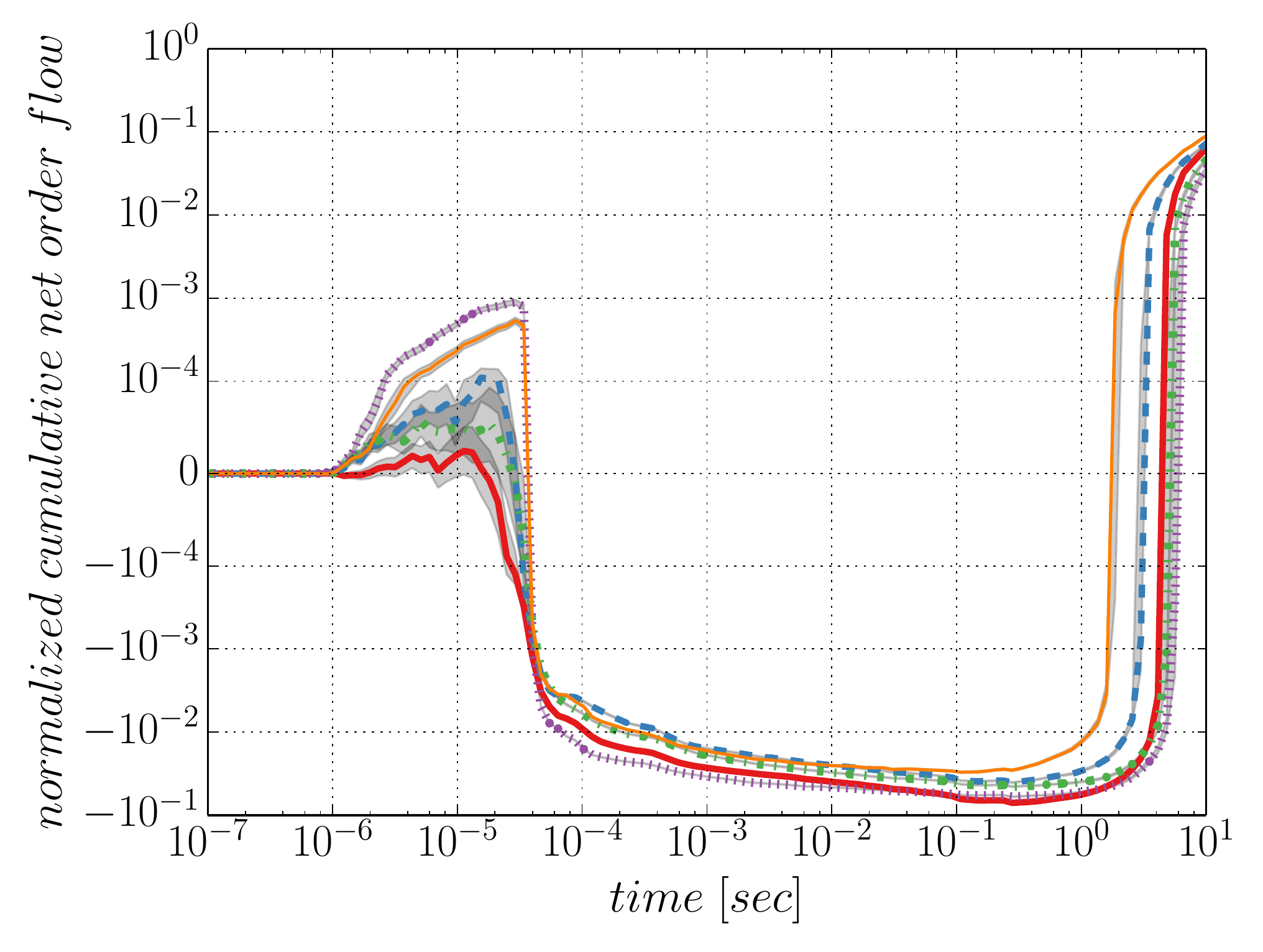}
  \caption{Mean cumulative net order flow for the same-side best queue $\bar V^{s}$ for (solid red) MSFT, (dashed blue) INTC, (dash-dotted green) CSCO, (dotted violet) YHOO, and (thin solid orange) MU, during the given times immediately after the arrival of a price-maintaining market order. Each stock's order flow is rescaled according to the mean number of shares at the best quotes (see Table \ref{tab1} and the description in the main text). The grey shaded region surrounding each curve indicates one standard error, which we estimate by calculating the sample standard deviation of the output at each lag, across $10000$ independent bootstrap samples of the data. In both panels, we plot the time $\tau$ in logarithmic coordinates. In the left panel, we plot our results with a linear scale on the vertical axis. In the right panel, we plot our results with a symmetric-logarithm scale on the vertical axis, with a linear region for $\left|\bar V^{s}\right| \leq 10^{-4}$ and a logarithmic region for $\left|\bar V^{s}\right|>10^{-4}$, to illustrate the behaviour for both positive and negative values with small magnitude.}
\label{fig:v_5stocks_after_same}
\end{figure}

Figure~\ref{fig:v_5stocks_after_same} shows the temporal evolution of the normalized $\bar V^{s}$ for each of the stocks in our sample. For all 5 stocks, we observe a sequence of 4 distinct order-flow phases: the first between the market order arrival and about $10^{-6}$ seconds, the second between about $10^{-6}$ seconds and about $10^{-4.5}$ seconds, the third between about $10^{-4.5}$ seconds and about $10^{-0.5}$ seconds, and the fourth after about $10^{-0.5}$ seconds. Despite the considerable differences in their trading activity (see Table \ref{tab1}), the progression  through these phases happens approximately contemporaneously for each of the stocks that we study. We now describe each of these phases in detail.

For all 5 stocks in our sample, the mean cumulative net order flow is exactly 0 until about $10^{-6}$ seconds after the market order arrival. Consistently with our results in Section \ref{subsec:separation}, this suggest that the Nasdaq order-matching server has a platform latency of about $10^{-6}$ seconds. 

After the platform-latency phase, and until about $10^{-4.5}$ seconds after the market order arrival, we observe a period during which net order flow is positive but very small. As we discuss in Section \ref{subsec:before}, the direction of this net order flow matches that of net order flow very shortly \emph{before} the market order arrival. We therefore conjecture that the small net order flow during this time period is actually a continuation of the same net order flow from before the market order arrival. In this way, we argue that this positive net order flow does not occur as a response to the market order arrival, but rather that it occurs in spite of it, because traders have not yet had the opportunity to react to the market order arrival. For this reason, we call this time period the \emph{response-latency phase},\footnote{\citet{Kirilenko:2015latency} argues that this type of latency consists of two components: market-feed latency, which is the time it takes for an automated trading platform to disseminate market data, and communication latency, which is the time it takes for a message to travel between a trader's computer and an automated trading platform.} because the limit orders processed by the server during this period are likely to have been submitted without knowledge of the previous market order arrival.

After the response-latency phase, net order flow suddenly becomes negative for all 5 stocks, and it remains negative until about $10^{-0.5}$ seconds after the market order arrival. It is extremely unlikely that such rapid order submissions could be achieved by a human trader, so we argue that this activity is generated by electronic trading algorithms responding to the market order arrival by cancelling previously submitted limit orders at the same-side best quote. We therefore call this time period the \emph{high-speed reaction phase}. We return to a further discussion of this point in Section \ref{sec:discussion}.

After about $10^{-0.5}$ seconds, we observe a \emph{lower-speed reaction phase,} during which the net order flow becomes positive as traders submit new limit orders at the best quotes. This net inflow of liquidity causes the queue to be restored to its initial (i.e., post-market order) length after about $10^{0.5} \approx 3$ seconds, then subsequently to exceed this length. This inflow of limit orders is consistent with the widely reported ``stimulated refill'' effect (see e.g., \citet{Bouchaud:2006random,Gerig:2007theory,Rosenau}), by which the arrival of a market order encourages other traders to submit new limit orders at the same price. We again return to a further discussion of this point in Section \ref{sec:discussion}.

\subsection{Net Order Flow at the Opposite-Side Best Quote}\label{subsec:after_opposite}

In this section, we consider the \emph{opposite-side mean net order flow}\begin{equation}\bar V^{o}(\tau) := \frac{1}{|\mathfrak M(\tau)|}\sum_{k \in \mathfrak M(\tau)} \left[ \mathbf 1_{\omega_k>0}W^A(t_k,\tau) + \mathbf 1_{\omega_k<0} W^{B}(t_k,\tau)\right]\end{equation}at a given time $\tau$ seconds after the arrival of a market order.\footnote{Similarly to Equation (\ref{eq:v_bar_same_after}), we use the indicator function $\mathbf{1}$ to align buy-side and sell-side activity (see Section \ref{subsec:after_same}).} As in Section \ref{subsec:after_same}, we normalize our results for each stock by dividing each stock's value of $\bar V^{o}(\tau)$ by the mean number of shares at the best quote for that stock, when averaged across our entire sample. We again use a standard non-parametric bootstrap to estimate the standard error of our estimates at each $\tau$.

\begin{figure}
\centering
	\includegraphics[width = 0.45\linewidth]{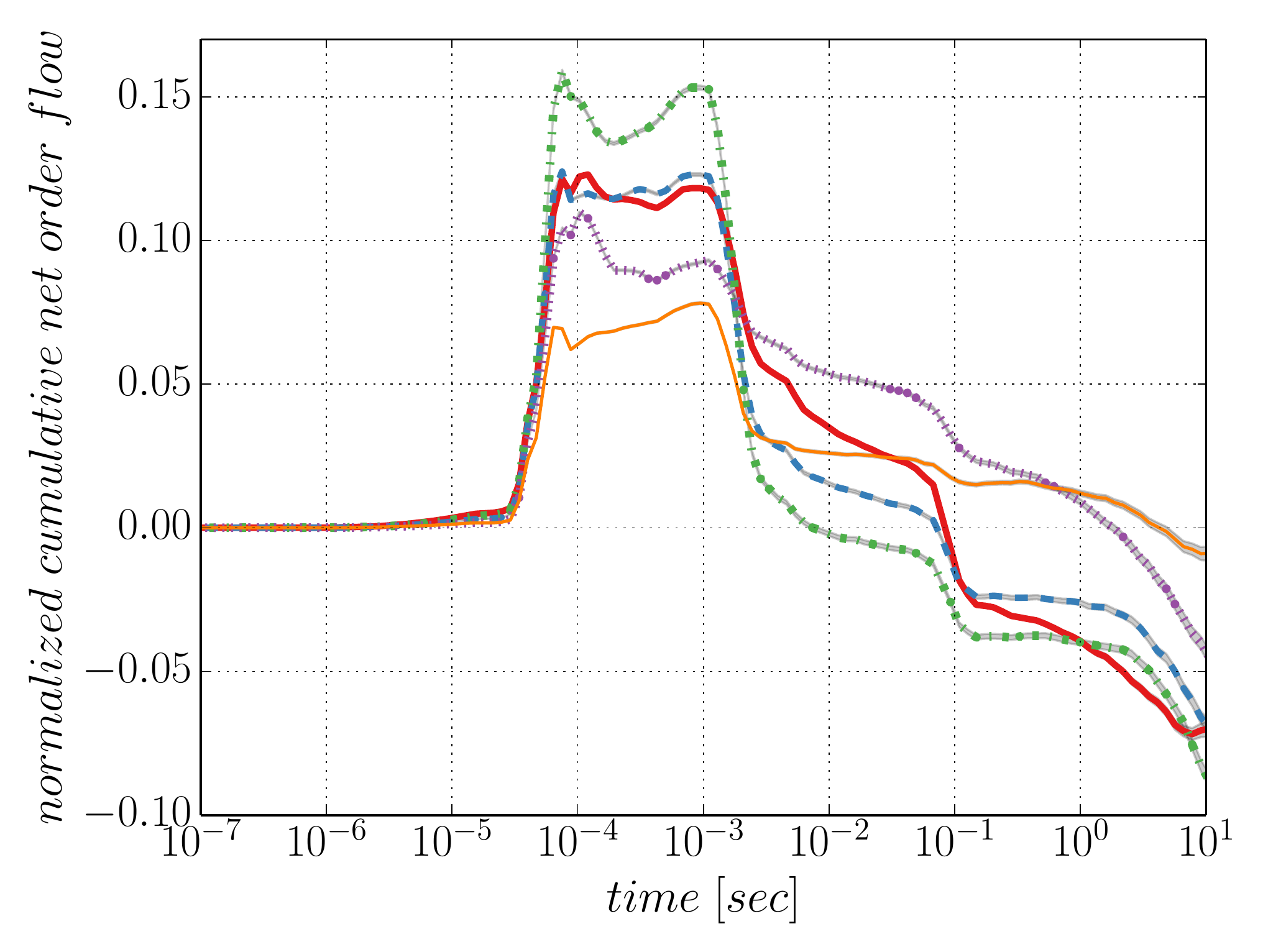}
  \includegraphics[width = 0.45\linewidth]{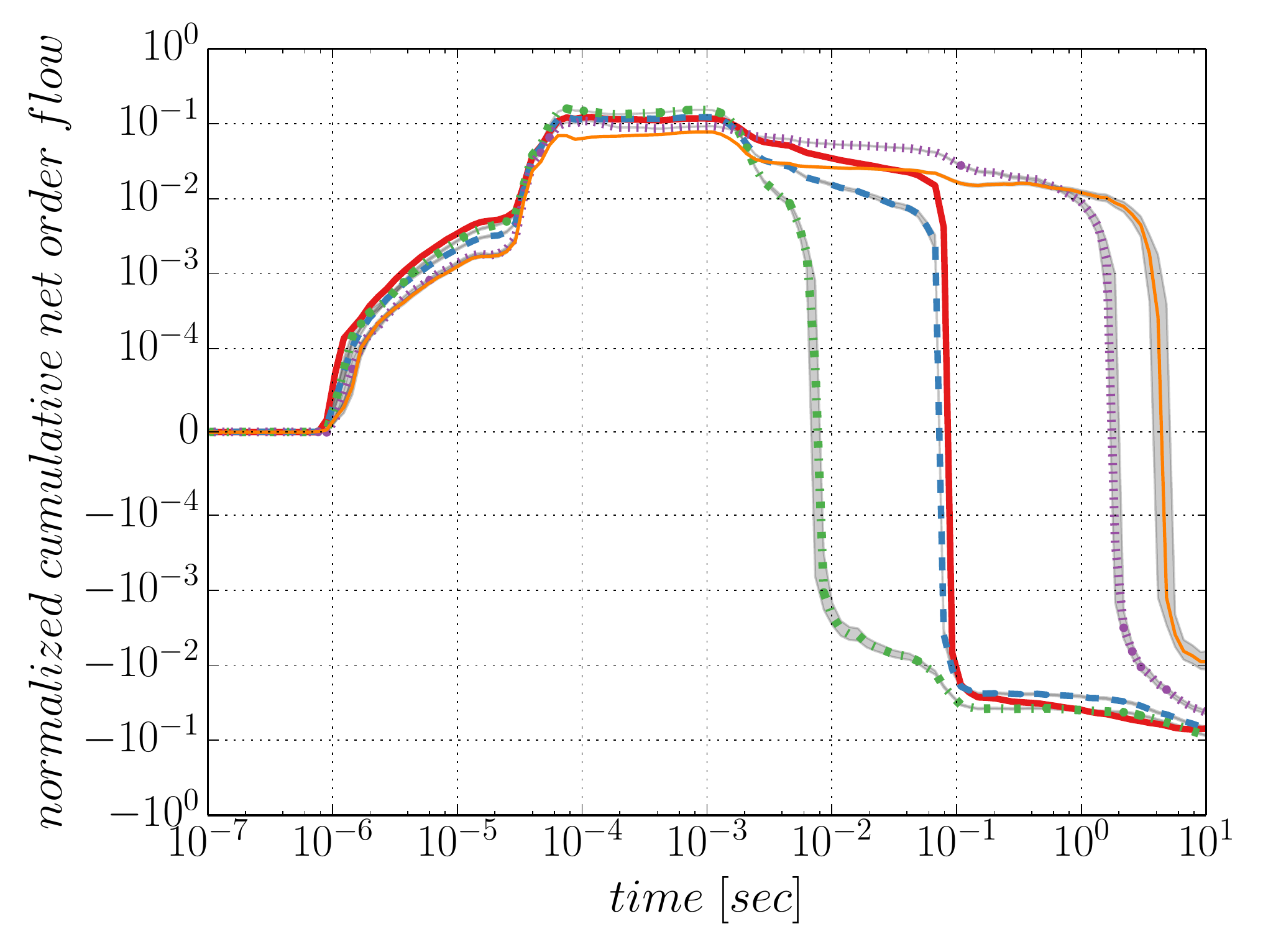}
  \caption{Mean cumulative net order flow for the opposite-side best queue $\bar V^{o}$ for (solid red) MSFT, (dashed blue) INTC, (dash-dotted green) CSCO, (dotted violet) YHOO, and (thin solid orange) MU, during the given times immediately after the arrival of a price-maintaining market order. Each stock's order flow is rescaled according to the mean number of shares at the best quotes (see Table \ref{tab1} and the description in the main text). The grey shaded region surrounding each curve indicates one standard error, which we estimate by calculating the sample standard deviation of the output at each lag, across $10000$ independent bootstrap samples of the data. In both panels, we plot the time $\tau$ in logarithmic coordinates. In the left panel, we plot our results with a linear scale on the vertical axis. In the right panel, we plot our results with a symmetric-logarithm scale on the vertical axis, with a linear region for $\left|\bar V^{o}\right| \leq 10^{-4}$ and a logarithmic region for $\left|\bar V^{o}\right|>10^{-4}$, to illustrate the behaviour for both positive and negative values with small magnitude.}
\label{fig:v_5stocks_after_opp}
\end{figure}

Figure~\ref{fig:v_5stocks_after_opp} shows the temporal evolution of the normalized $\bar V^{o}$ for each of the stocks in our sample. We again observe a sequence of 4 distinct order-flow phases. Similarly to our results for the same-side best quote (see Section \ref{subsec:after_same}), each stock's transition between the first and second phase occurs at about $10^{-6}$ seconds, and each stock's transition between the second and the third phase occurs at about $10^{-4.5}$ seconds. In contrast to our results in Section \ref{subsec:after_same}, the transition time between the third and fourth phase varies across the different stocks in our sample. We now describe each of these phases in detail.

Similarly to the same-side activity, the mean cumulative net order flow at the opposite-side best quote is exactly 0 until about $10^{-6}$ seconds after the market order arrival. This result is consistent with our hypothesis that this period is a system-latency phase that occurs due to system latency in the Nasdaq order-matching server.

Also similarly to the same-side activity, the cumulative net order flow at the opposite-side best quote is positive between about $10^{-6}$ seconds and about $10^{-4.5}$ seconds after the market order arrival. Again, this result is consistent with our hypothesis that this positive net order flow constitutes a response-latency phase that occurs because traders have not yet had the opportunity to react to the market order arrival (i.e., that the limit orders processed by the server during this period are likely to have been submitted without knowledge of the previous market order arrival).

After the response-latency phase, the cumulative mean net order flow at the opposite-side best queue increases sharply, then remains approximately constant until about $10^{-3}$ seconds after the market order arrival. As we argued in Section \ref{subsec:after_same}, the extremely fast response times during this high-frequency reaction phase suggest that this order flow is generated by electronic trading algorithms, which in this case submit new limit orders at the opposite-side best quote. This activity is consistent with a decreased fear of adverse selection at the opposite-side best queue after the market order arrival.

After the high-frequency reaction phase, we again observe a lower-speed reaction phase, during which the mean net flow of orders becomes negative because many limit orders are cancelled. This outflow of limit orders continues until several seconds after the market order arrival. This observation is consistent with the hypothesis that liquidity providers consider the expected waiting costs of remaining in a limit order queue, which increase as the queue becomes longer. We return to a further discussion of this point in Section \ref{sec:discussion}.

\subsection{Net Order Flow Before a Market Order Arrival}\label{subsec:before}

In Section~\ref{subsec:before}, we investigated the mean cumulative net order flow after a market order arrival. In this section, we calculate the corresponding statistics before the arrival of a market order by considering $\bar V^{s}(\tau)$ and $\bar V^{o}(\tau)$ for negative values of $\tau$. In this way, we condition on the direction of the (price-maintaining) market order arrival at time $t_{i+1}$, and count backwards in time from this arrival.

\begin{figure}
\centering
	\includegraphics[width = 0.45\linewidth]{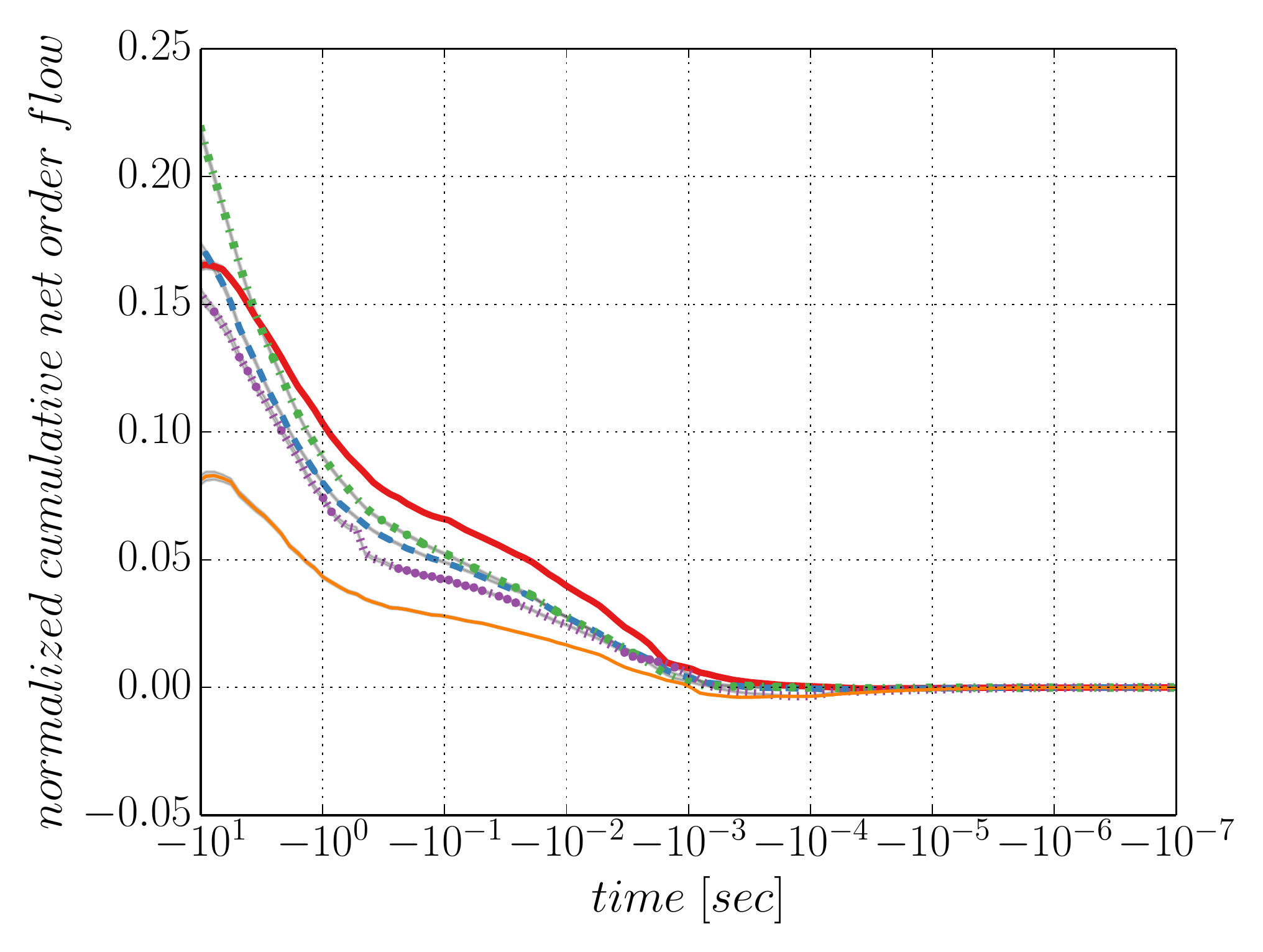}
  \includegraphics[width = 0.45\linewidth]{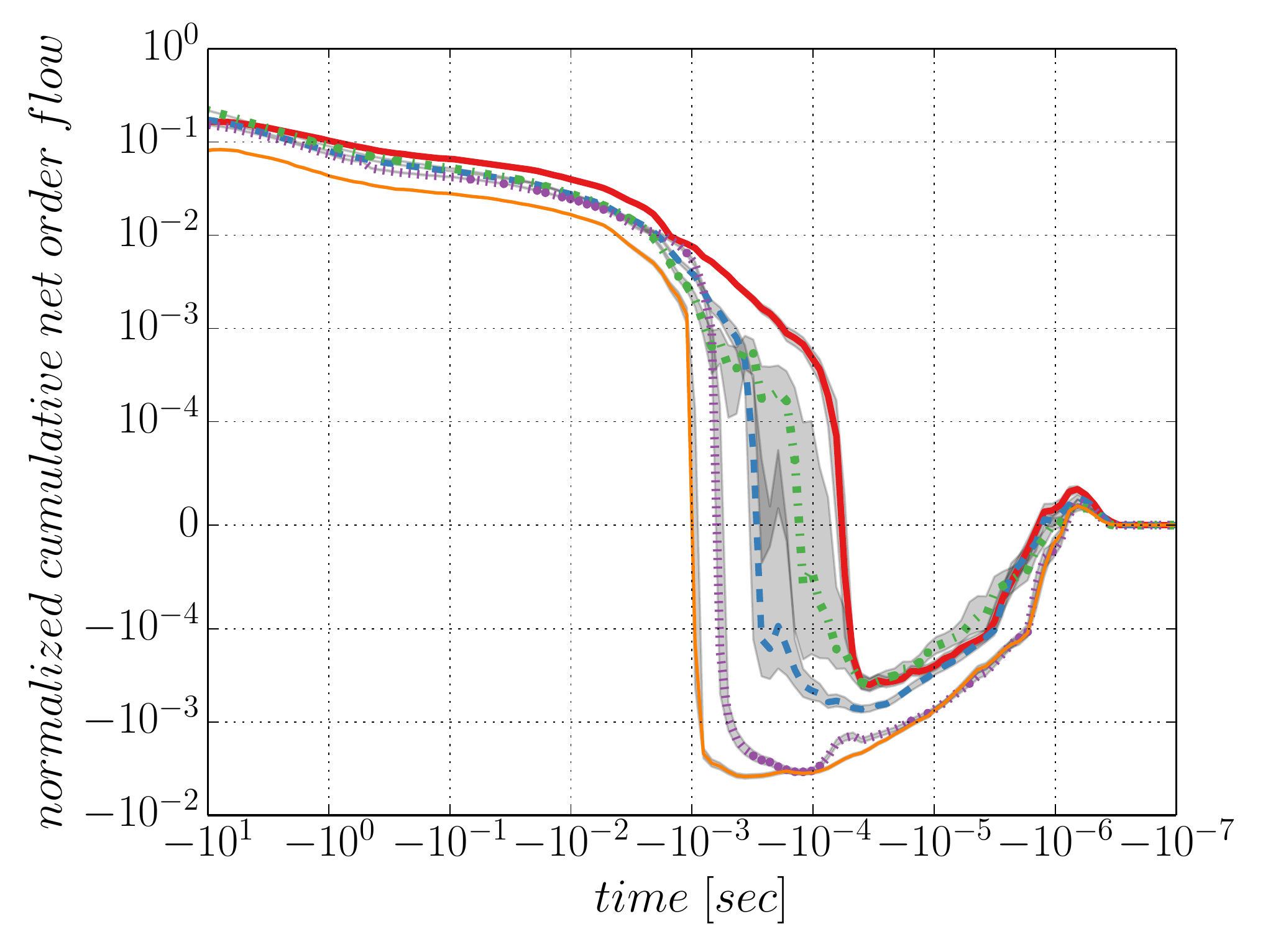}
	  \caption{Mean cumulative net order flow for the same-side best queue $\bar V^{s}$ for (solid red) MSFT, (dashed blue) INTC, (dash-dotted green) CSCO, (dotted violet) YHOO, and (thin solid orange) MU, during the given times immediately before the arrival of a price-maintaining market order. Each stock's order flow is rescaled according to the mean number of shares at the best quotes (see Table \ref{tab1} and the description in the main text). The grey shaded region surrounding each curve indicates one standard error, which we estimate by calculating the sample standard deviation of the output at each lag, across $10000$ independent bootstrap samples of the data. In both panels, we plot the time $\tau$ in logarithmic coordinates. In the left panel, we plot our results with a linear scale on the vertical axis. In the right panel, we plot our results with a symmetric-logarithm scale on the vertical axis, with a linear region for $\left|\bar V^{s}\right| \leq 10^{-4}$ and a logarithmic region for $\left|\bar V^{s}\right|>10^{-4}$, to illustrate the behaviour for both positive and negative values with small magnitude.}
\label{fig:v_5stocks_before_same}
\end{figure}

Figure~\ref{fig:v_5stocks_before_same} shows the the temporal evolution of $\bar V^{s}$ for $\tau<0$. The same-side best queue shrinks in the period leading up to the market order arrival (which occurs at $\tau=0$), which indicates that many traders cancel their existing limit orders at the same-side best quote. Shortly before the market order arrival, however, the mean net order flow reverses direction. The time at which this reversal occurs varies somewhat across the different stocks that we study, but this is unsurprising given that traders do not know when (or even whether) the upcoming market order will arrive, so the synchronicity between their behaviour is less strong than we observed subsequent to the market order arrival (when the arrival time is known). This positive net order flow continues until just before --- and, as we discussed in Section \ref{subsec:after_same}, shortly after --- the market order arrival. Immediately preceding the market order, we again observe a short system-latency time with similar magnitude to the one that we observe after the market order arrival.

\begin{figure}
\centering
	\includegraphics[width = 0.45\linewidth]{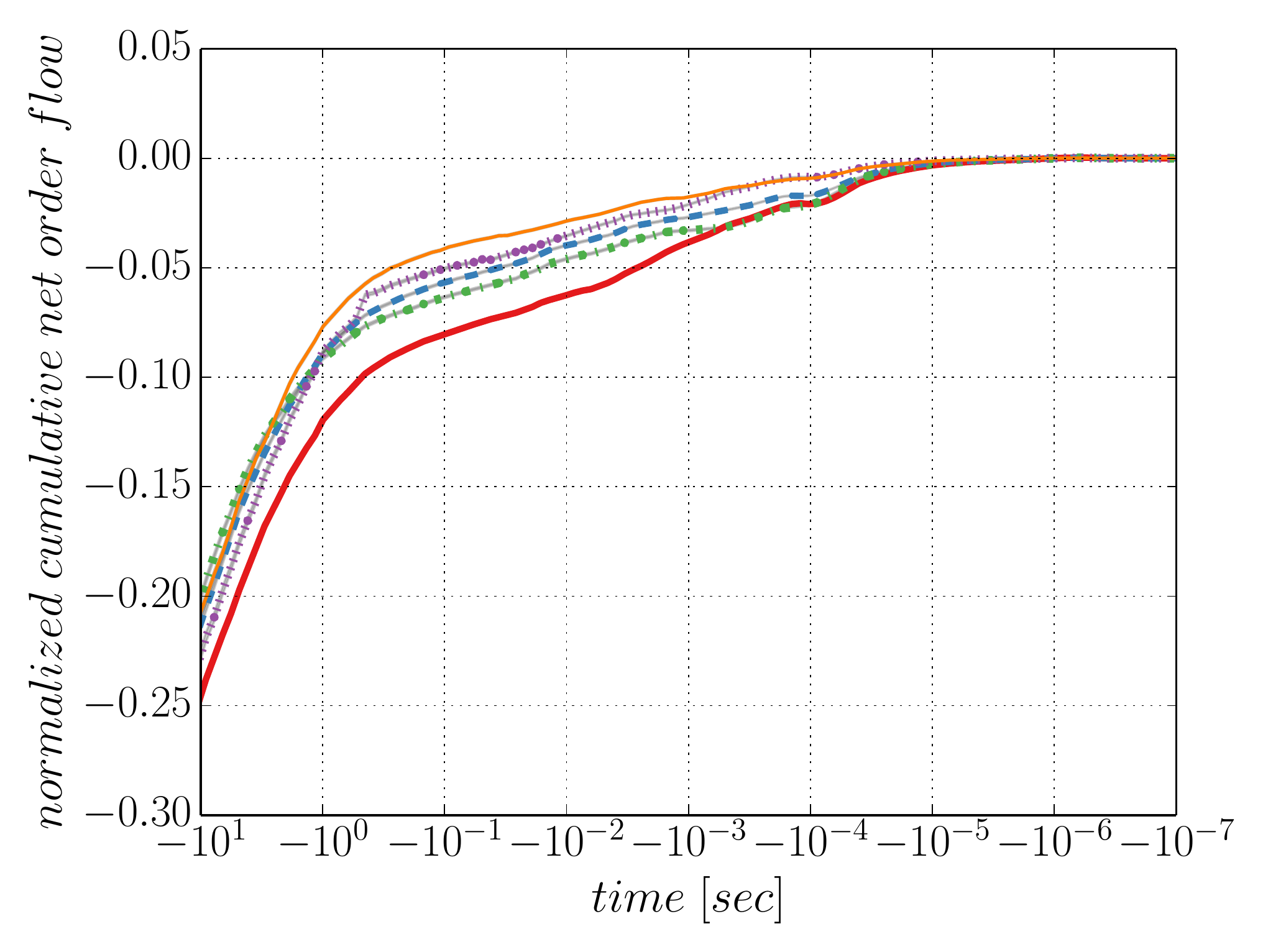}
  \includegraphics[width = 0.45\linewidth]{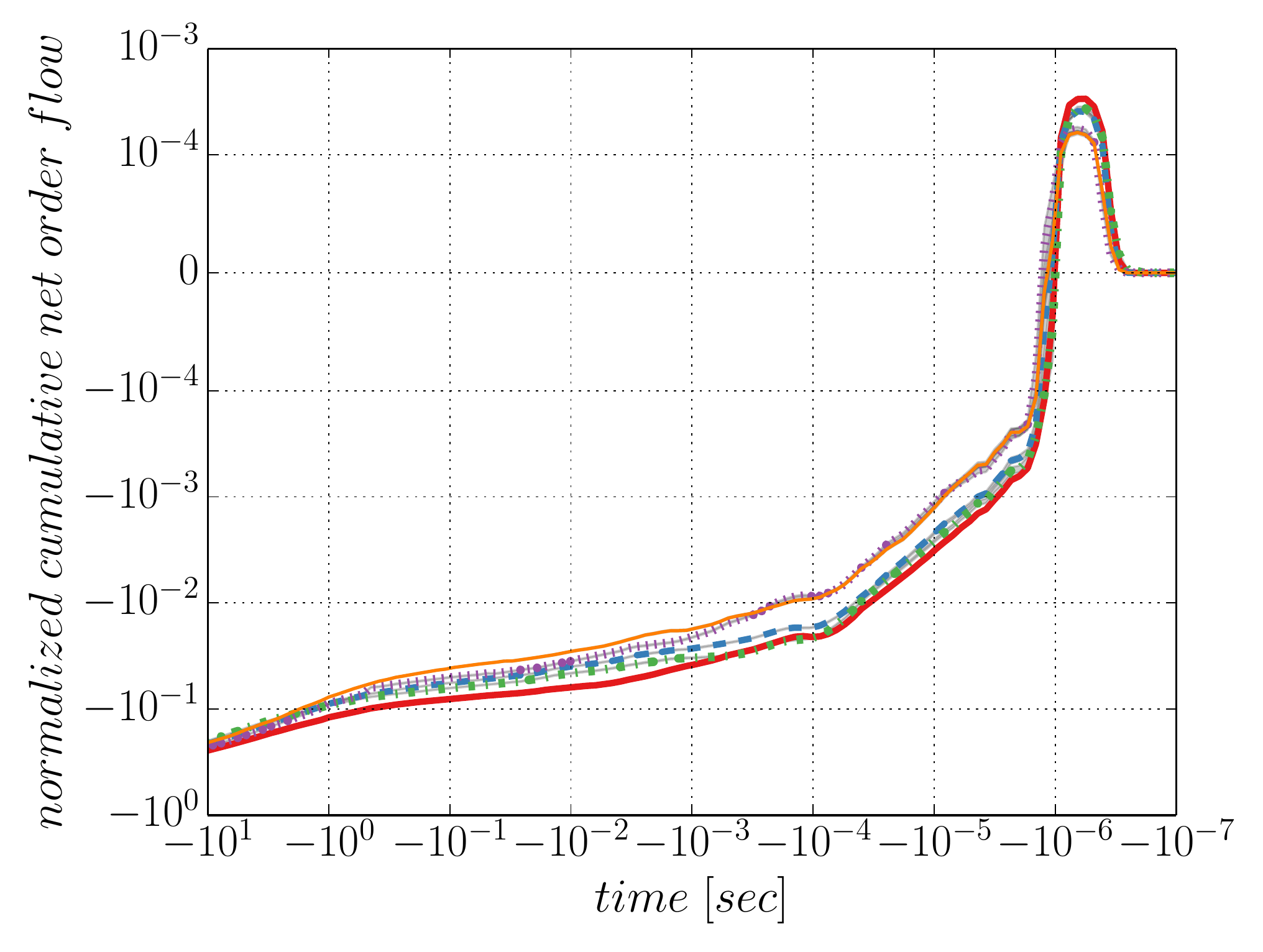}
	  \caption{Mean cumulative net order flow for the opposite-side best queue $\bar V^{o}$ for (solid red) MSFT, (dashed blue) INTC, (dash-dotted green) CSCO, (dotted violet) YHOO, and (thin solid orange) MU, during the given times immediately before the arrival of a price-maintaining market order. Each stock's order flow is rescaled according to the mean number of shares at the best quotes (see Table \ref{tab1} and the description in the main text). The grey shaded region surrounding each curve indicates one standard error, which we estimate by calculating the sample standard deviation of the output at each lag, across $10000$ independent bootstrap samples of the data. In both panels, we plot the time $\tau$ in logarithmic coordinates. In the left panel, we plot our results with a linear scale on the vertical axis. In the right panel, we plot our results with a symmetric-logarithm scale on the vertical axis, with a linear region for $\left|\bar V^{o}\right| \leq 10^{-4}$ and a logarithmic region for $\left|\bar V^{o}\right|>10^{-4}$, to illustrate the behaviour for both positive and negative values with small magnitude.}
\label{fig:v_5stocks_before_opp}
\end{figure}

Figure~\ref{fig:v_5stocks_before_opp} shows the the temporal evolution of $\bar V^{o}$ (i.e., the opposite-side best quote) for negative values of $\tau$. In contrast to the same-side activity, the length of the opposite-side best queue increases on average during the lead-up to the market order arrival. This finding is consistent with the findings of some other empirical studies of order flow \citep{Avellaneda:2011forecasting,Gould:2015imbalance}, which have noted how market order arrivals occur more often when there is a strong imbalance (i.e., normalized difference) between the lengths of the bid and ask queues. Our results in Figures~\ref{fig:v_5stocks_before_same} and \ref{fig:v_5stocks_before_opp} are consistent with this hypothesis, because they illustrate that the imbalance between the same-side and opposite-side queues increases during the period immediately before the market order arrival. We return to this discussion in Section \ref{sec:discussion}.

\subsection{Price Movements}\label{subsec:dynamic_price}

In this section, we repeat our experiments from Sections \ref{subsec:after_same}, \ref{subsec:after_opposite} and \ref{subsec:before}, but when relaxing the restriction that the quote prices $b_t$ and $a_t$ must remain constant during the period that we study. More precisely, we calculate the cumulative net order flows at $b_t$ and $a_t$ as follows:

\begin{itemize}
\item For limit order flow that does not change the value of $b_t$ and $a_t$, we use the same measurements as in Sections \ref{subsec:after_same}, \ref{subsec:after_opposite} and \ref{subsec:before}.
\item For limit order flow that causes the order queue at $b_t$ or $a_t$ to deplete to zero (and therefore causes $b_t$ to decrease or $a_t$ to increase), we measure the final order departure from the old price then continue to monitor order flow at the new values of $b_t$ and $a_t$.
\item For limit order flow that arrives inside the spread (and therefore causes $b_t$ to increase or $a_t$ to decrease), we measure this limit order arrival and continue to monitor subsequent limit order flow at the new values of $b_t$ and $a_t$.
\end{itemize}

\begin{figure}
\centering
	\includegraphics[width = 0.49\linewidth]{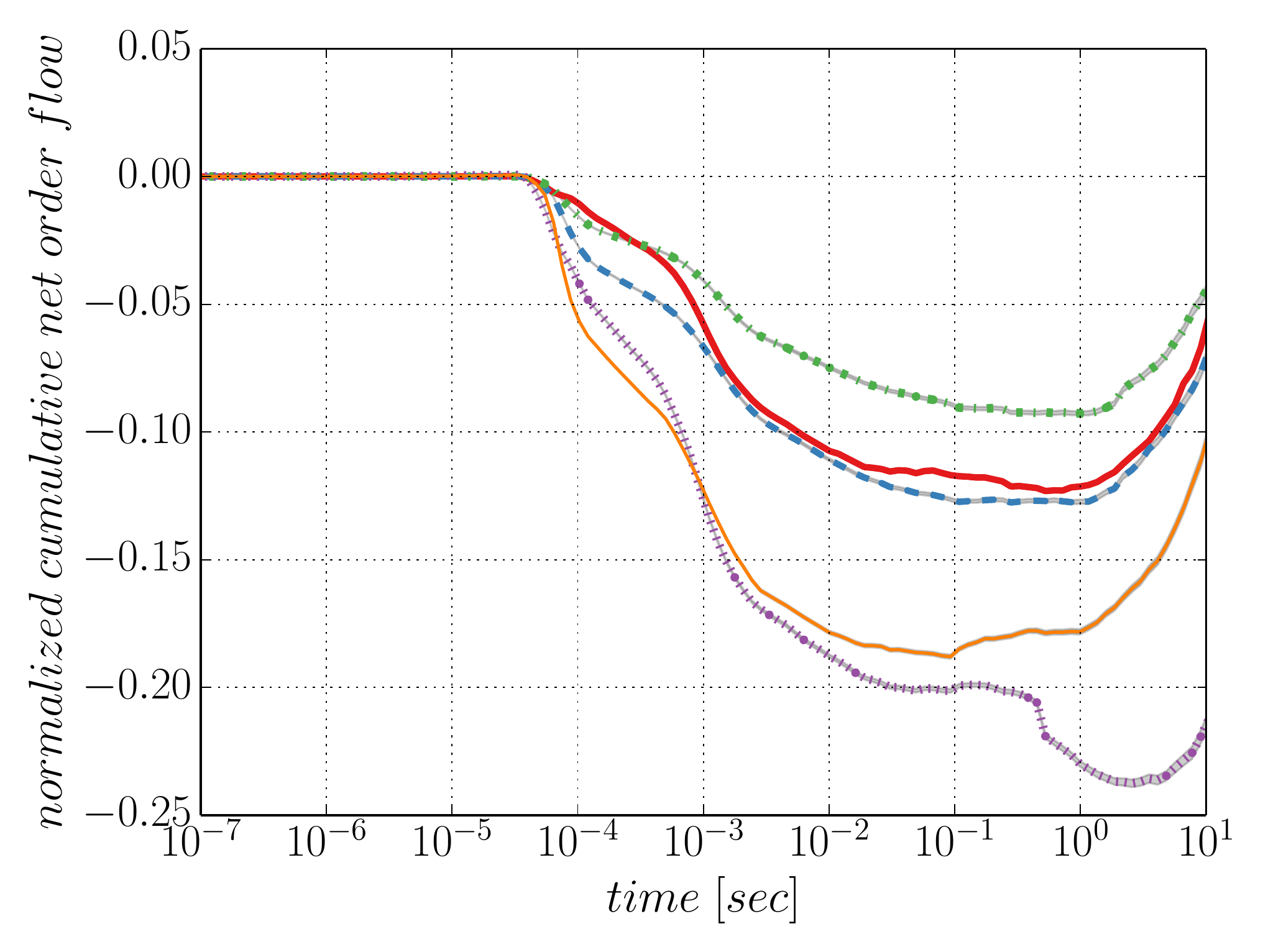}
	\includegraphics[width = 0.49\linewidth]{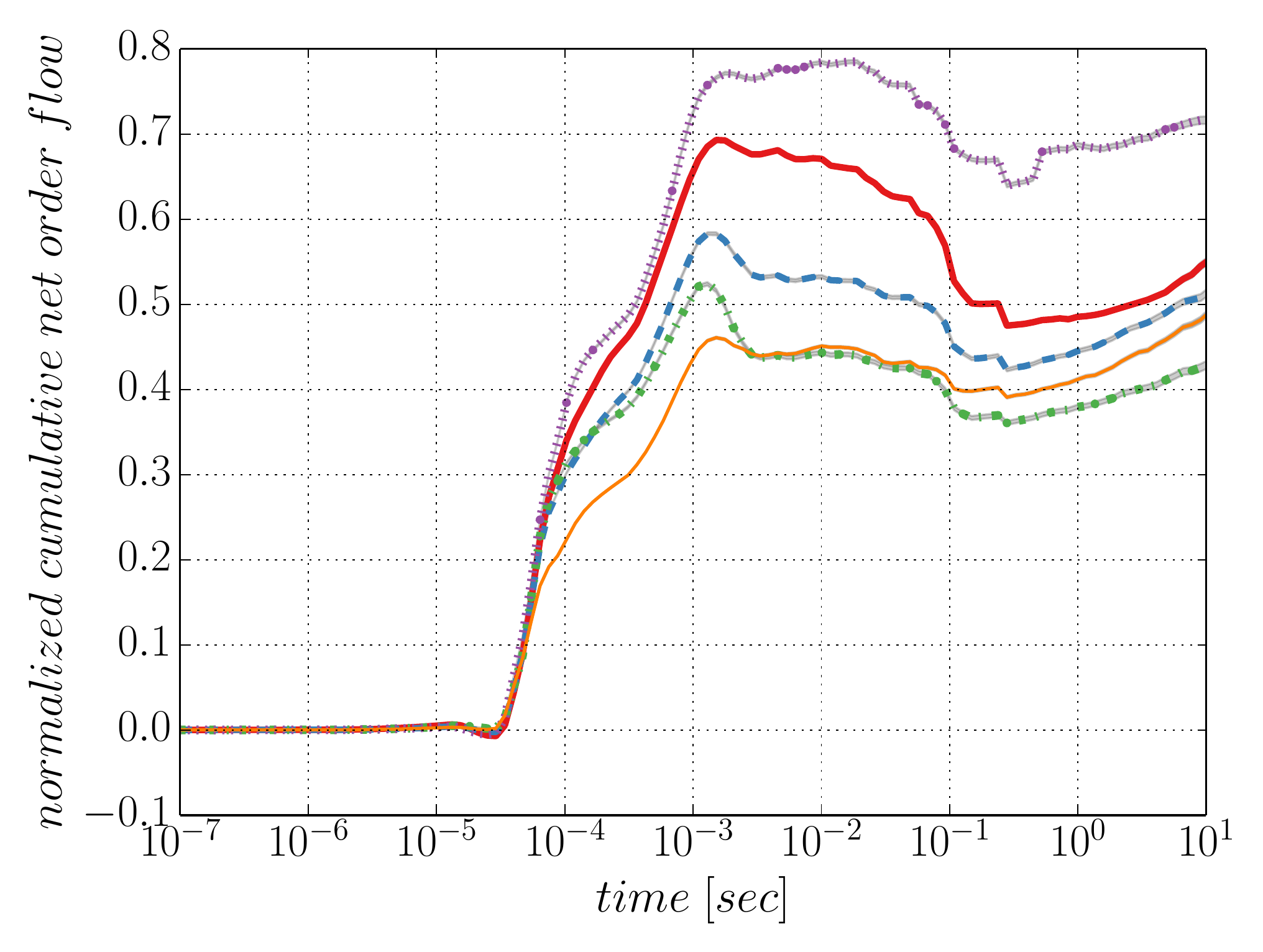}
	\includegraphics[width = 0.49\linewidth]{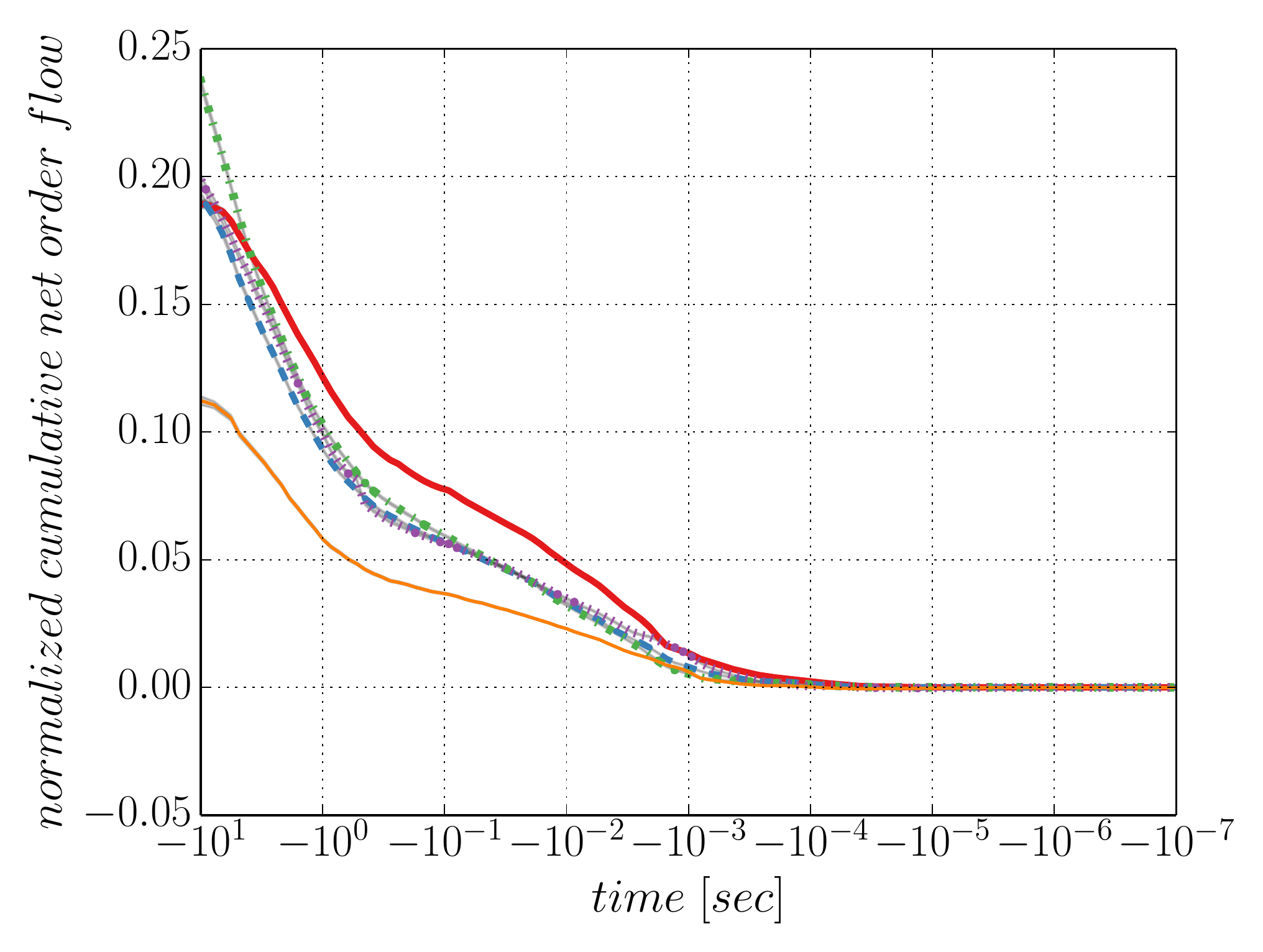}
	\includegraphics[width = 0.49\linewidth]{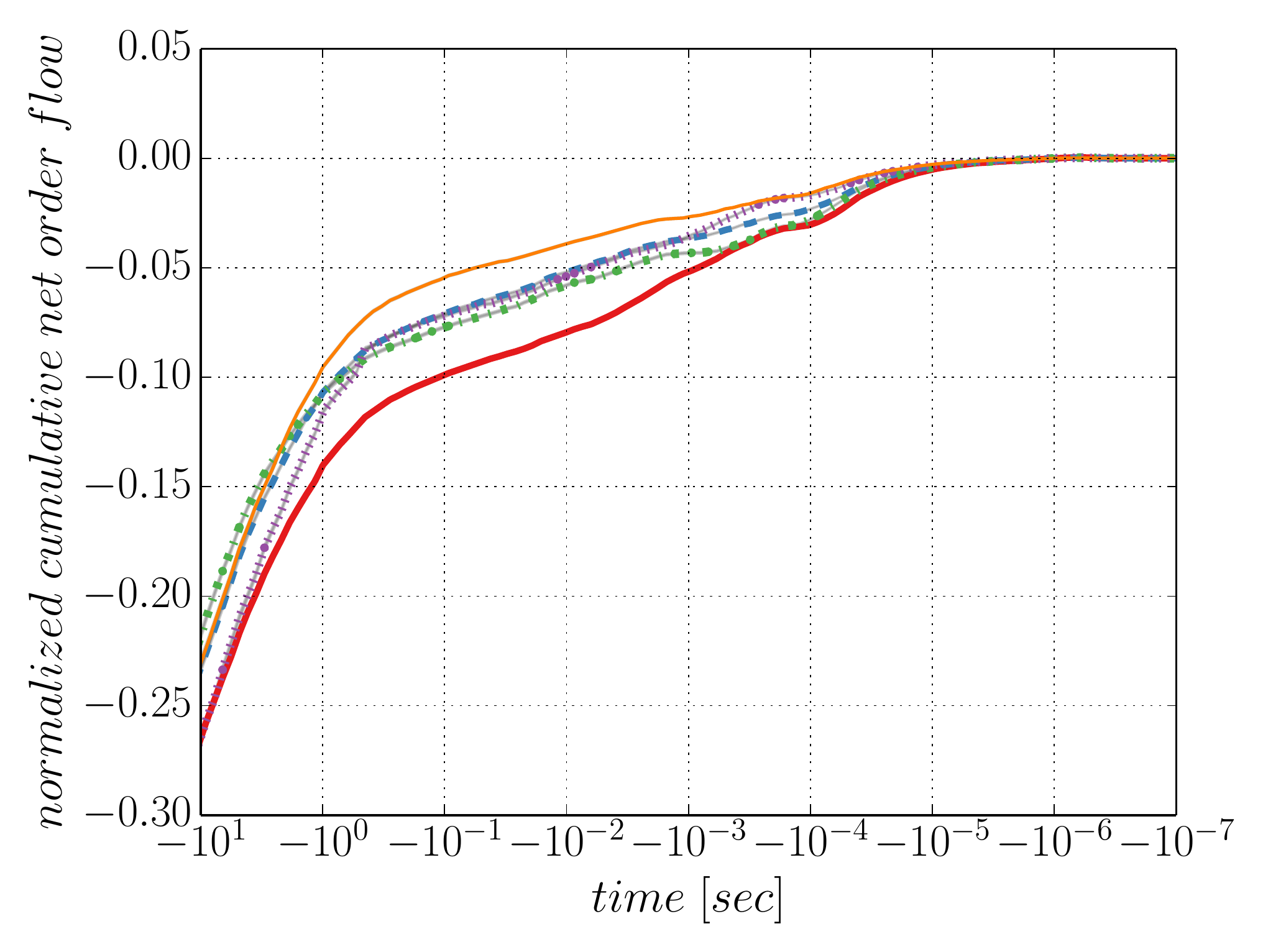}	
\caption{Mean cumulative net order flow for the (left column) same-side best queue $\bar V^{s}$ and (right column) opposite-side best queue $\bar{V}^o$, during the period (top row) after and (bottom row) before the arrival of a market order for (solid red) MSFT, (dashed blue) INTC, (dash-dotted green) CSCO, (dotted violet) YHOO, and (thin solid orange) MU. Each stock's order flow is rescaled according to the mean number of shares at the best quotes (see Table \ref{tab1}). The grey shaded region surrounding each curve indicates one standard error, which we estimate by calculating the sample standard deviation of the output at each lag, across $10000$ independent bootstrap samples of the data.}
\label{fig:afterbefore_sameopp}
\end{figure}

Figure~\ref{fig:afterbefore_sameopp} shows the temporal evolution of the mean cumulative net order flow at the (left column) same-side and (right column) opposite-side best quotes (top row) after and (bottom row) the arrival of a $T$-separated market order. Because we now consider order flow at $b_t$ and $a_t$, irrespective of their prices, we no longer demand that the original market order is price-maintaining.

In all 4 panels of Figure~\ref{fig:afterbefore_sameopp}, the qualitative shapes of the mean cumulative net order flow trajectories are similar to those that we observed in Sections \ref{subsec:after_same}, \ref{subsec:after_opposite}, and \ref{subsec:before}, for which we only considered order flow that did not change the values of $b_t$ or $a_t$. When considering activity before the market order arrival, we also find that the magnitudes of the cumulative net order flow are similar to those that we observed in Section \ref{subsec:before}. However, when considering activity after the market order arrival, the magnitudes of the cumulative net order flows are much larger in Figure~\ref{fig:afterbefore_sameopp} than for the corresponding figures in Sections \ref{subsec:after_same} and \ref{subsec:after_opposite}. We return to the discussion of this interesting result in Section \ref{sec:discussion}.

\section{Discussion}\label{sec:discussion}

Our empirical results raise many interesting points for discussion. In this section, we address these points, propose some possible explanations for the behaviour that we observe, and highlight some interesting avenues for future research.

We first address the issue of latency. As we discuss in Section \ref{subsec:after_same}, our results suggest that the total latency time (i.e., the minimum time between a market order arrival and the corresponding reactions from other market participants) consists of two phases: a platform-latency phase, which lasts about $10^{-6}$ seconds, and a response-latency phase, which lasts about $10^{-4.5}$ seconds. In a recent study of electronic trading, \citet{Kirilenko:2015latency} reported that platform latency on the Bolsa de Valores, Mercadorias \& Futuros de Sao Paulo exchange during 2014 varied across several orders of magnitude, ranging from hundreds of microseconds to tens of milliseconds. The shortest platform-latency times that we observe on Nasdaq during 2015 (see Figures \ref{fig:Deltat_lower} and \ref{fig:v_5stocks_after_same}) are shorter than the shortest platform-latency times observed by \citet{Kirilenko:2015latency}. However, because we are not able to measure the full distribution of latency times in our data, we are not able to discern whether some institutions experience much longer platform-latency times. Further investigation into the variability of platform-latency times on Nasdaq would be an interesting topic for future research.

As we argue in Section \ref{subsec:after_same}, the cumulative net order flow that we observe between about $10^{-6}$ seconds and about $10^{-4.5}$ seconds after a market arrival (see Figure \ref{fig:v_5stocks_after_same}) is consistent with the existence of a response-latency phase, which \citet{Kirilenko:2015latency} argues consists of both market-feed latency and communication latency. The sharp change in cumulative net order flow that occurs after this period suggest that the shortest total latency times achieved by the fastest traders on the platform are about $10^{-4.5}$ seconds.

\begin{figure}
  \centering
  \includegraphics[width = 0.45\linewidth]{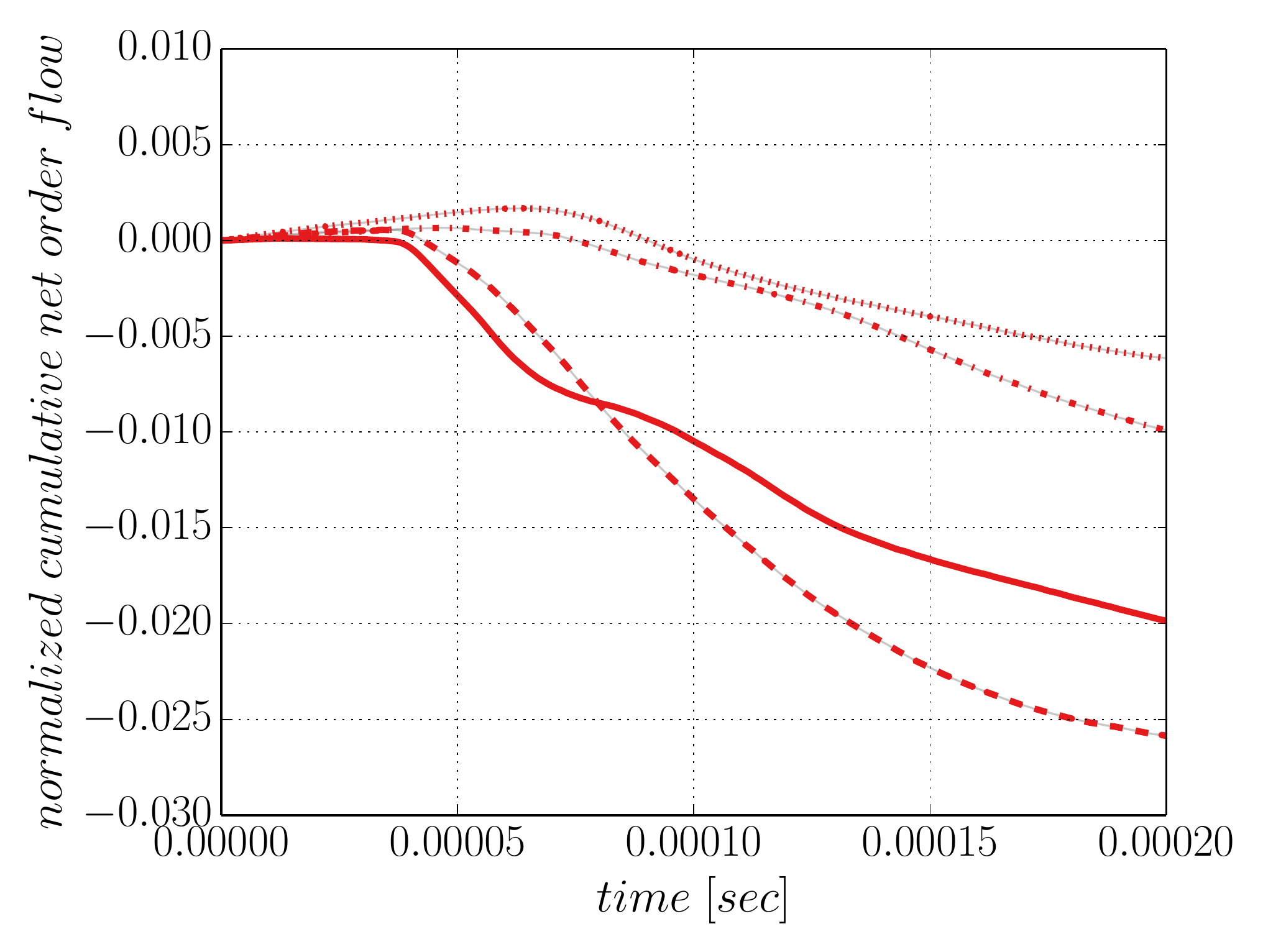}
  \includegraphics[width = 0.45\linewidth]{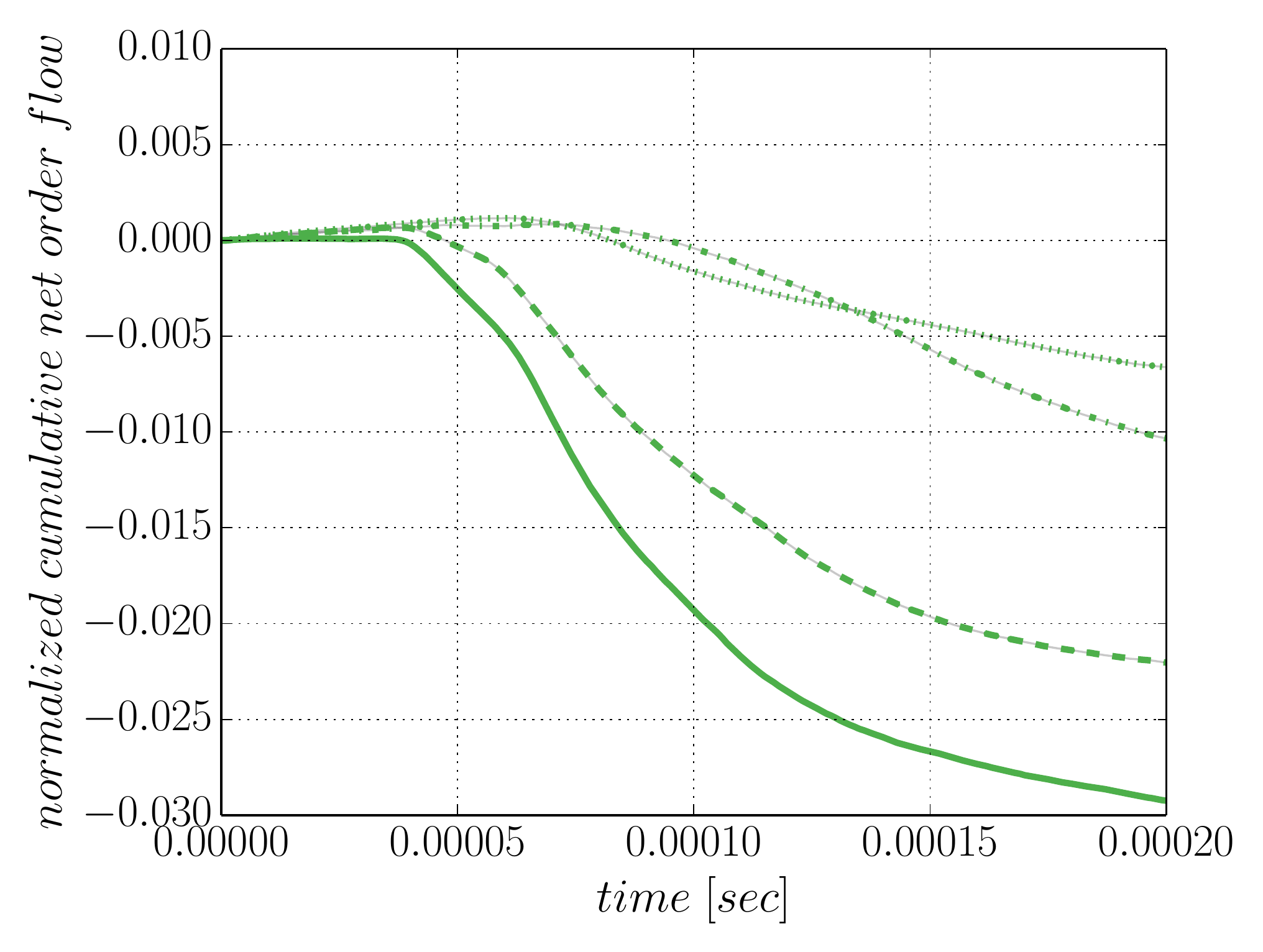}
  \caption{Mean cumulative net order flow for the same-side best queue $\bar V^{s}$ for (left panel) MSFT and (right panel) CSCO during the given times immediately after the arrival of a price-maintaining market order, in the years (dotted curves) 2012, (dash-dotted curves) 2013, (dashed curves) 2014, and (solid curves) 2015. Each stock's order flow is rescaled according to the mean number of shares at the best quotes in the respective year.}
  \label{fig:differentyears}
\end{figure}

Understanding the total latency time is a key consideration for high-frequency traders, who seek to submit orders extremely quickly to respond rapidly to changes in market state. Due to technological advances in both computer processors and  telecommunications networks, it seems reasonable to expect that the total latency times experienced by high-frequency traders has fallen considerably over time. Indeed, we are able to observe this effect directly on Nasdaq by repeating our experiments using older data from the platform. Figure \ref{fig:differentyears} shows the mean cumulative net order flow for the same-side best queue $\bar V^{s}$ for MSFT and CSCO during the given times immediately after the arrival of a price-maintaining market order, in the years 2012, 2013, 2014, and 2015. In each case, the time $\tau$ corresponding to the first local maximum of the curves, which we argue corresponds to the initial market reaction to the market order, decreases with each subsequent year. Moreover, the location of each year's local maximum is approximately the same for MSFT as it is for CSCO, so we argue that this local maximum corresponds to the total latency time in the given year. Therefore, these results suggest that the total latency time decreased in each subsequent year during this period.

After the latency period, we observe a period of strong net outflow until about $10^{-0.5}$ seconds after the market order arrival, followed by a period of strong net inflow (see Figure \ref{fig:v_5stocks_after_same}). Several other studies have addressed this strong net inflow (see e.g., \citet{Bouchaud:2006random,Gerig:2007theory,Rosenau}), and have argued that this phenomenon is caused by stimulated refill, by which the market order arrival encourages other traders to submit new limit orders at the same price. This behaviour is consistent with the hypothesis that when deciding how to act, some traders consider the expected waiting cost of remaining in a limit order queue, which becomes shorter after the market order arrival. However, this story does not provide an explanation for the preceding period of strong net outflow, during which many traders cancel their existing limit orders.

In the context of stimulated refill, these cancellations are surprising, because they suggest that some traders cancel their limit orders despite their newly increased priority in the limit order queue. Why would traders cancel these orders in this situation? We conjecture that the answer to this question lies in liquidity providers' --- and, given the fast reaction times, particularly high-frequency traders' --- increased fear of adverse selection. Specifically, if a liquidity provider observes the arrival of a buy (respectively, sell) market order, he/she may fear that the market order's owner has private information to suggest that the current value of $a_t$ is too low (respectively, $b_t$ is too high), and that the owner of the market order conducted a trade to ``pick off'' one or more mispriced limit orders. Moreover, because each market order arrival shortens the length of the same-side best queue, traders with an existing limit order that remains in this queue after the trade are exposed to an increased likelihood that the next arriving market order will consume all the limit orders at this price, and will therefore generate immediate price impact.

\begin{figure}
  \centering
  \includegraphics[width = 0.45\linewidth]{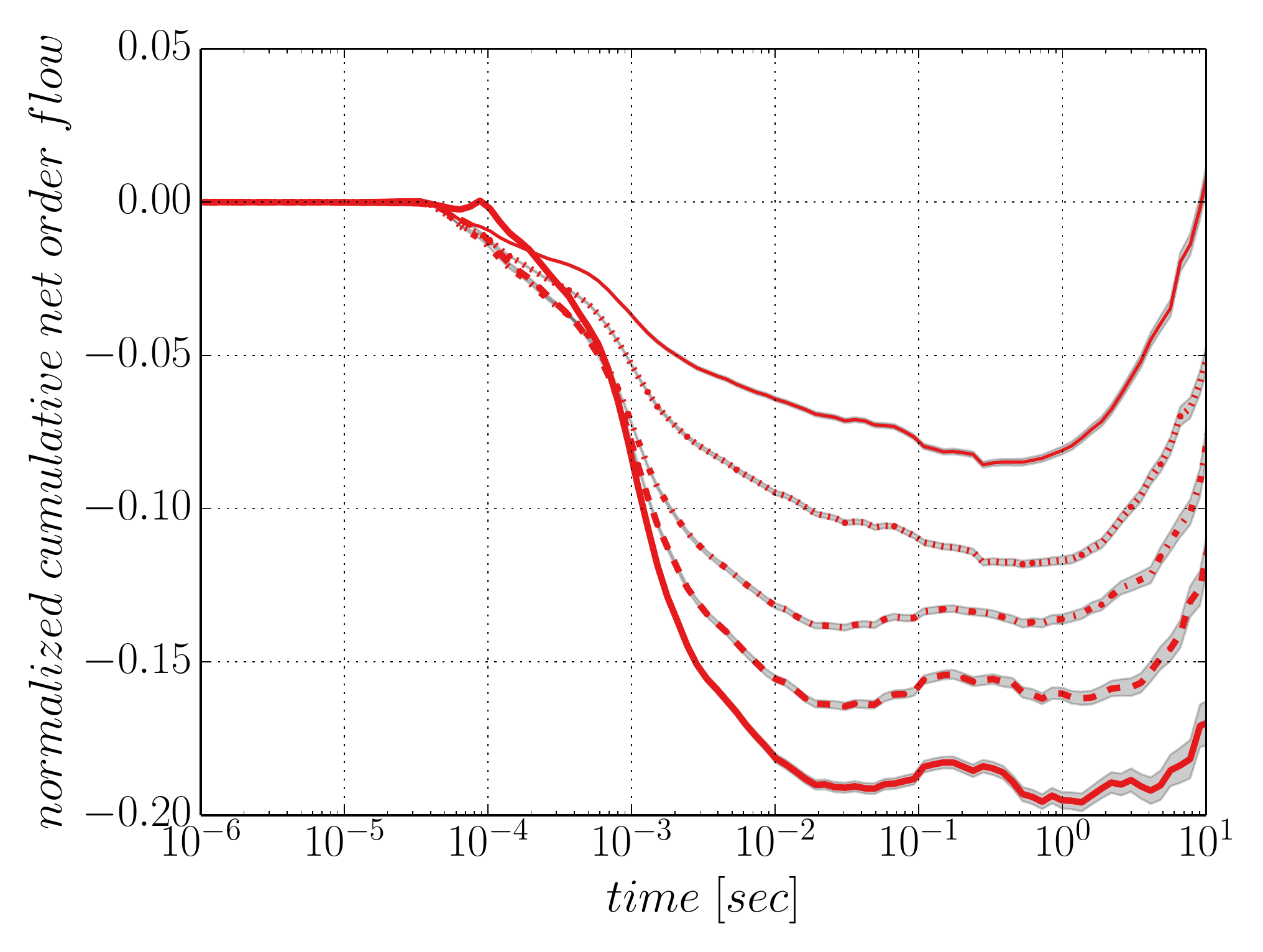}
  \includegraphics[width = 0.45\linewidth]{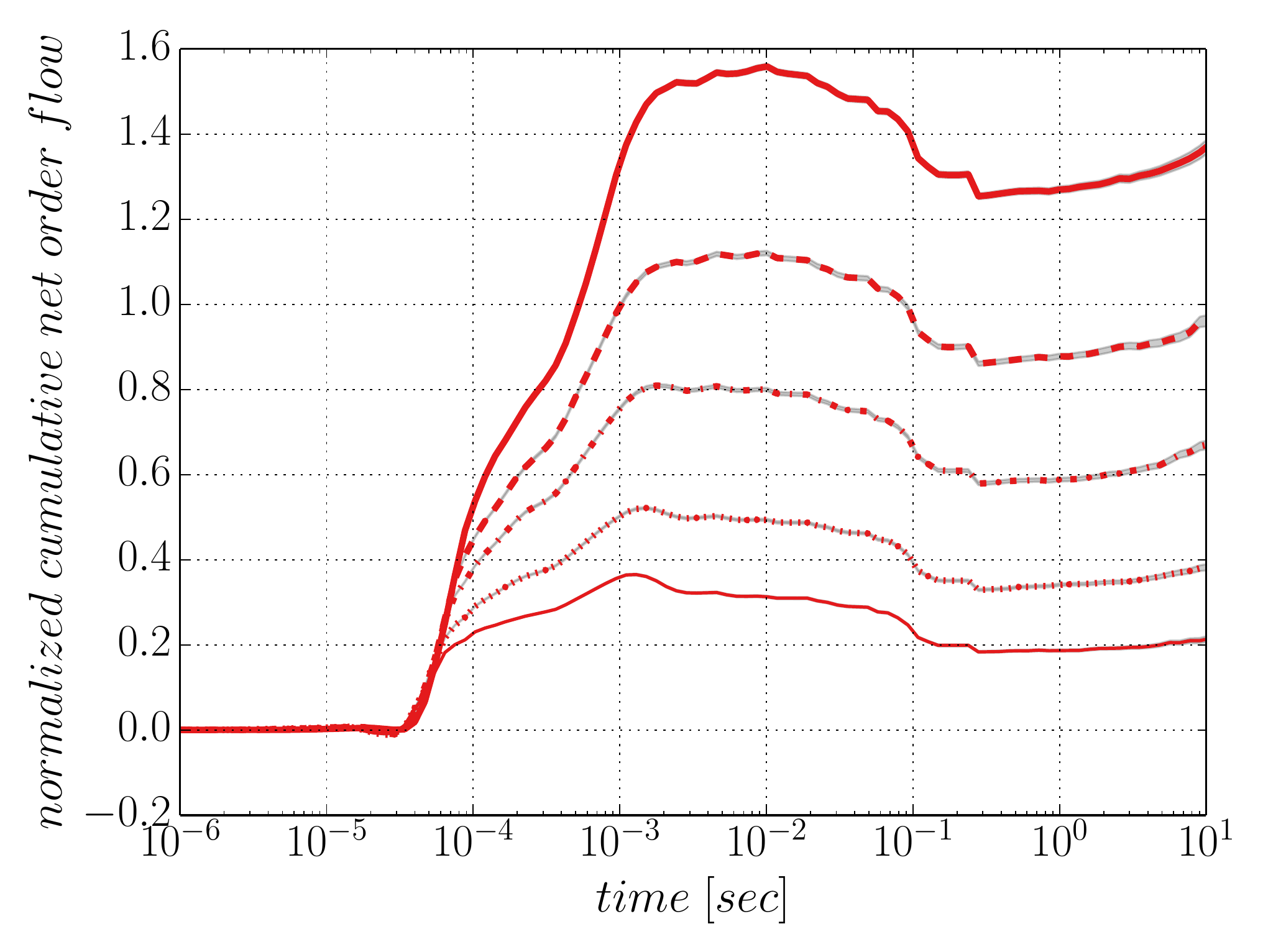}
  \caption{Normalized mean cumulative net order flow for the (left) same-side best queue $\bar V^{s}$ and (right) opposite-side best queue $\bar{V}^o$ for MSFT during the given times immediately after the arrival of a market order with size in the (thin solid curve) first, (dotted curve) second, (dash-dotted curve) third, (dashed curve) fourth, and (thick solid curve) fifth quintiles of the empirical market order size distribution. The grey shaded region surrounding each curve indicates one standard error, which we estimate by calculating the sample standard deviation of the output at each lag, across $10000$ independent bootstrap samples of the data.}
  \label{fig:differentvolumes}
\end{figure}

To examine the plausibility of this explanation, we also repeat our analysis of the cumulative net order flow at the same-side best quotes, but when partitioning our observations according to the size of the arriving market order. Specifically, we partition all market order sizes into 5 bins containing an approximately equal number of data points, and we calculate the mean cumulative net order flow $\bar V^{s}$ among the trajectories in the first, second, third, fourth, and fifth quintiles separately. We plot these results in Figure \ref{fig:differentvolumes}. The plots clearly illustrate that the net outflow of limit orders is stronger after larger market order arrivals. This result is consistent with our hypothesis that this net outflow is due to traders' fear of adverse selection, because larger market orders could be interpreted as stronger signals of private information.

Our results at the opposite-side best quote (see Section \ref{subsec:after_opposite}) also provide interesting insight into the possible motivations for traders' behaviour. Shortly after the arrival of a market order, we observe net order flows that are consistent with the same platform-latency and response-latency periods that we observe at the same-side best quote. After about $10^{-4.5}$ seconds, we then observe a strong net inflow of limit orders at the opposite-side best quote. This strong net inflow is consistent with our hypothesis that some traders regard the arrival of a buy (respectively, sell) market order to be a signal that the asset's price is likely to rise (respectively, fall), thus encouraging these traders to submit new buy limit orders at $b_t$ (respectively, sell limit orders at $a_t$).

After this rapid net inflow of limit orders at the opposite-side best quote, we then observe a gradual net outflow. We propose that this behaviour occurs because some traders regard the expected waiting costs associated with the (recently lengthened) limit order queue to have become unattractive, and may therefore cancel their orders. It would be interesting to test this hypothesis by performing an empirical study of where in the limit order queue cancellations occur most frequently. If our previous interpretation is correct, then we would expect cancellations to occur more frequently among limit orders later in the queue,\footnote{\citet{Gareche:2013fokker} provides a brief remark that this is indeed the case on Nasdaq.} because these limit orders experience higher expected waiting costs. Investigating this question more deeply is an interesting avenue for future research.

The behaviour that we observe before the arrival of a market order (see Section \ref{subsec:before}) also raises several interesting points for discussion. Shortly before the market order arrival, we again observe a short period with 0 net order flow (see Figures \ref{fig:v_5stocks_before_same} and \ref{fig:v_5stocks_before_opp}), which we argue occurs due to platform latency. Interestingly, the smallest platform-latency times that we observe in these plots are slightly shorter than the smallest platform-latency times that we observe after the arrival of a market order (see Figures \ref{fig:v_5stocks_after_same} and \ref{fig:v_5stocks_after_opp}).

We propose two possible explanations for why this might occur. First, the platform-latency time associated with a limit order arrival or cancellation may be shorter than the platform-latency time associated with a market order arrival. Therefore, the platform-latency time for the final event before a market order arrival may be shorter than the platform-latency time for the first event after a market order arrival. Second, the clock used to record time stamps for market order arrivals may be different to the clock used to record time stamps for limit order arrivals or cancellations.\footnote{This is indeed the case on many other trading platforms \citep{Hautsch:2011econometrics}.} If these clocks are slightly mis-aligned, then the apparent platform-latency time before a market order arrival may be slightly different to the apparent platform-latency time after a market order arrival.

Shortly before the arrival of a market order, we see a small but sudden cancellation of orders. This effect is particularly apparent at the opposite-side best quote. Why should this be so? We conjecture that this phenomenon occurs because some traders who submit market orders do so immediately cancelling their own limit orders. We propose two possible reasons why a trader might act in this way. First, consider a trader who wishes to buy a given quantity of an asset within a given time interval. To seek a favorable price for the trade, the trader may choose to first submit a buy limit order at $b_t$ and wait to see whether this order becomes matched. If so, the trader has received a better price for the trade than he/she would have achieved by submitting a market order at that start of the time interval. If not, then the trader may opt to cancel this buy limit order and instead submit a buy market order to complete the necessary purchase. Adopting this simple strategy would enable a trader to avoid excessive waiting costs while preserving the possibility of gaining a better execution price if his/her limit order becomes matched before the end of the time interval.

Second, some traders on electronic trading platforms implement so-called ``spoofing'' strategies \citep{Lee:2013microstructure}, which involve the rapid submission and cancellation of orders to entice other market participants to behave in a certain way or to mislead them about the true state of the LOB. The rapid cancellations that we observe before the arrival of a market order could be consistent with some traders implementing spoofing strategies. At present, relatively little is known about the possible consequences of spoofing, so a more detailed analysis of this question would be an interesting avenue for future research.

Before these rapid cancellations occur, the same-side best queue gradually shortens (see Figure \ref{fig:v_5stocks_before_same}) while the opposite-side best queue gradually lengthens (see Figure \ref{fig:v_5stocks_before_opp}). This results suggests that, on average, the \emph{LOB imbalance} (i.e., the normalized difference between the queue lengths at the same-side and opposite-side best quotes) gradually strengthens during this period. Several recent empirical studies have reported strong statistical links between LOB imbalance and subsequent order flow (see, e.g., \citet{Avellaneda:2011forecasting,Gould:2015imbalance}). Our results provide two possible explanations to help explain why this might occur. The first possible explanation is that liquidity takers who seek to buy (respectively, sell) the asset are more likely to do so by submitting a market order when they observe the ask (respectively, bid) queue almost depleted, to avoid the possibility that the queue will empty before they are able to trade. This strategy is often called selective liquidity taking. In our empirical calculations, we only consider market order arrivals that do not fully deplete the best queue at their time of arrival. Therefore, we do not expect that selective liquidity taking strongly impacts our results, because selective liquidity takers would likely consume the whole queue with their market order. Therefore, we do not regard this possible explanation to be the primary cause of the behaviour that we observe.

The second possible explanation, which we regard as much more plausible, is that some traders use the LOB imbalance as a predictor of future market activity. For example, some traders may predict that an asset's price is likely to increase whenever its LOB imbalance exceeds a certain threshold, and may therefore submit a market order to attempt to profit from this situation. Such traders may act not because they believe that the same-side best queue will fully deplete imminently, but rather because they seek to take advantage of some form of order-flow ``momentum'' or to improve their execution strategy. Despite the appealing simplicity of this explanation, it does not address how this LOB imbalance emerges and evolves in the first place. We seek to address this interesting question in a future publication.

Comparing the results that we obtain when we condition on the quote prices remaining constant (see Sections \ref{subsec:after_same},  \ref{subsec:after_opposite}, and \ref{subsec:before}) to those when we allow the quote prices to change (see Section \ref{subsec:dynamic_price}) reveals interesting insight into the way that traders submit and cancel orders across different prices. After the arrival of a market order, we observe a much stronger net inflow at the opposite-side best quote when we allow the quote prices to move (see Figure \ref{fig:afterbefore_sameopp}) than when we condition on the quote prices remaining constant (see Figure \ref{fig:v_5stocks_after_opp}). We conjecture that this effect occurs due to some traders submitting new buy (respectively, sell) limit orders inside the spread after the arrival of a buy (respectively, sell) market order. By definition, the arrival of a buy (respectively, sell) limit order inside the spread causes $b_t$ to increase (respectively, $a_t$ to decrease). Therefore, the arrival of any such order would be observed in a trajectory in which we allow the quote price to move, but not when we condition on the quote prices remaining constant. In this way, conditioning on quote prices remaining constant can be regarded as a censored sample of the full net inflow of orders.

Similarly, we observe a much stronger net outflow at the same-side best quote when we allow the quote prices to move (see Figure \ref{fig:afterbefore_sameopp}) than when we condition on the quote prices remaining constant (see Figure \ref{fig:v_5stocks_after_same}). In this case, conditioning on quote prices remaining constant can be regarded as censoring the sample of the true net outflow of orders, because this conditioning does not reveal the subsequent outflow of orders at other prices after the best queue depletes to 0.

Interestingly, net order flow at the same-side and opposite-side best quotes before the arrival of a market order is largely unaffected by whether or not we condition on the quote prices remaining constant during this period. We find this result rather surprising, because it brings into question how heavily market orders really are correlated with the limit order flow preceding their arrival. It seems reasonable to assume that a new limit order arrival inside the bid--ask spread would stimulate new market order arrivals, and therefore cause different statistical properties to emerge in Figures~\ref{fig:v_5stocks_before_same} and \ref{fig:v_5stocks_before_opp} than in Figure~\ref{fig:afterbefore_sameopp}. However, this does not appear to be the case. Therefore, it seems that a new limit order arriving inside the spread causes a similar influence on subsequent market order arrivals to a new limit order arriving at the previous best quote.

As a final point for discussion, we address the similarities and differences that we observe between the mean cumulative net order flows for the different stocks in our sample. After the arrival of a market order, the times at which the stocks undergo transitions between the different order-flow phases are remarkably similar for each of the stocks that we study (see Figures \ref{fig:v_5stocks_after_same} and \ref{fig:v_5stocks_after_opp}). Before the arrival of a market order, the same synchronicity holds at the opposite-side best quote (see Figure \ref{fig:v_5stocks_before_opp}), but is less strong at the same-side best quote (see Figure \ref{fig:v_5stocks_before_same}), which we conjecture is due to the uncertainty surrounding when (or whether) the upcoming market order will arrive. Similar results also arise when we allow the quote prices to change during the period of study (see Figure \ref{fig:afterbefore_sameopp}).

Despite uncovering these strong temporal similarities, our results illustrate that even after normalizing each stock's order flow according to its mean queue length (which, as we argue in Section \ref{subsec:after_same}, is a proxy for the stock's liquidity and activity), we still observe considerable quantitative differences across the cumulative net order flow for the different stocks in our sample. Although our simple normalization goes some way to reducing the cross-stock variation that we observe, it provides far from a perfect curve collapse. As part of our empirical analysis, we have also investigated a wide range of other possible normalizations based on intuitive physical properties of order flow and LOB state, such as the mean market order size and the total absolute order flow, but we have not been able to uncover a simple normalization that causes our the mean cumulative net order flows for the different stocks to collapse onto a single, universal curve. Therefore, we argue that if such a normalization exists, it is likely to consist of a nonlinear combination of several such factors, or of other factors entirely. Such a normalization would be an extremely useful tool, because it would help to provide insight into how the many interacting features of the system generate the complex order flows that we observe, and could therefore serve as a strong motivation for designing new LOB models. We therefore argue that this is a particularly interesting avenue for future research.

\section{Conclusions}\label{sec:conclusion}

In this paper, we have performed an empirical analysis of order flow in an LOB before and after the arrival of a market order. Thanks to the extremely detailed time resolution of our data, we were able to detect not only the widely reported phenomenon of stimulated refill, but also other other, more subtle effects that have not been reported elsewhere in the literature. We also studied and measured the impact of both platform latency and response latency, which are important considerations for high-frequency traders.

Our results show that limit order flows are strongly influenced by the arrivals of market orders. We highlighted that the LOB queue dynamics that we observe arise from the complex interplay between many different strategic considerations, and we provided several possible strategic motivations for these actions. Our results suggest that both expected waiting costs and the perceived risk of adverse selection play an important role in LOB dynamics.

One of the fundamental changes catalyzed by the widespread uptake of electronic trading is the blurring of lines between liquidity providers and liquidity takers. Historically, these roles were performed by different types of market participants, but in modern markets many traders both offer and consume liquidity according to their trading desires at a given moment. Therefore, the phenomena that we have observed should not be regarded as the consequences of only a specialized group of liquidity providers submitting limit orders to an LOB platform. Instead, our results illustrate that complex and often surprising phenomena can emerge from the interactions between the many different types of financial institutions that together comprise the diverse trading ecosystem in modern financial markets.

\paragraph*{Acknowledgements}

Julius Bonart thanks the Institute for Pure and Applied Mathematics at UCLA for hosting him as a visitor while part of this research was conducted. We thank Jean-Philippe Bouchaud, Rama Cont, Jonathan Donier, and Charles-Albert Lehalle for useful discussions. We thank Jonas Haase and Ruihong Huang for technical support. Julius Bonart gratefully acknowledges support from CFM and Martin D. Gould gratefully acknowledges support from the James S. McDonnell Foundation.

\bibliographystyle{abbrvnat}
\bibliography{bibli2.bib}

\begin{thebibliography}{33}
\providecommand{\natexlab}[1]{#1}
\providecommand{\url}[1]{\texttt{#1}}
\expandafter\ifx\csname urlstyle\endcsname\relax
  \providecommand{\doi}[1]{doi: #1}\else
  \providecommand{\doi}{doi: \begingroup \urlstyle{rm}\Url}\fi

\bibitem[Almgren and Chriss(2001)]{AlmgrenChriss}
R.~Almgren and N.~Chriss.
\newblock Optimal execution of portfolio transactions.
\newblock \emph{Journal of Risk}, 3:\penalty0 40, 2001.

\bibitem[Avellaneda et~al.(2011)Avellaneda, Reed, and
  Stoikov]{Avellaneda:2011forecasting}
M.~Avellaneda, J.~Reed, and S.~Stoikov.
\newblock Forecasting prices from level-{I} quotes in the presence of hidden
  liquidity.
\newblock \emph{Algorithmic Finance}, 1\penalty0 (1):\penalty0 35--43, 2011.

\bibitem[Bertsimas and Lo(1998)]{BertsimasLo}
D.~Bertsimas and A.~W. Lo.
\newblock Optimal control of execution costs.
\newblock \emph{Journal of Financial Markets}, 1:\penalty0 1--50, 1998.

\bibitem[Bouchaud et~al.(2006)Bouchaud, Kockelkoren, and
  Potters]{Bouchaud:2006random}
J.~P. Bouchaud, J.~Kockelkoren, and M.~Potters.
\newblock Random walks, liquidity molasses and critical response in financial
  markets.
\newblock \emph{Quantitative Finance}, 6\penalty0 (2):\penalty0 115--123, 2006.

\bibitem[Bouchaud et~al.(2009)Bouchaud, Farmer, and Lillo]{Bouchaud:2009digest}
J.~P. Bouchaud, J.~D. Farmer, and F.~Lillo.
\newblock How markets slowly digest changes in supply and demand.
\newblock In T.~Hens and K.~R. Schenk-Hopp{\'{e}}, editors, \emph{Handbook of
  Financial Markets: Dynamics and Evolution}, pages 57--160. North--Holland,
  Amsterdam, The Netherlands, 2009.

\bibitem[Cartea et~al.(2015)Cartea, Donnelly, and Jaimungal]{Donnelly}
A.~Cartea, R.~F. Donnelly, and S.~Jaimungal.
\newblock Enhanced trading strategies with order book signals.
\newblock \url{http://papers.ssrn.com/sol3/papers.cfm?abstract_id=2668277},
  2015.

\bibitem[Chakraborti et~al.(2011{\natexlab{a}})Chakraborti, Toke, Patriarca,
  and Abergel]{Chakraborti:2011agent}
A.~Chakraborti, I.~M. Toke, M.~Patriarca, and F.~Abergel.
\newblock Econophysics review {II}: {A}gent-based models.
\newblock \emph{Quantitative Finance}, 11\penalty0 (7):\penalty0 1013--1041,
  2011{\natexlab{a}}.

\bibitem[Chakraborti et~al.(2011{\natexlab{b}})Chakraborti, Toke, Patriarca,
  and Abergel]{Chakraborti:2011empirical}
A.~Chakraborti, I.~M. Toke, M.~Patriarca, and F.~Abergel.
\newblock Econophysics review {I}: {E}mpirical facts.
\newblock \emph{Quantitative Finance}, 11\penalty0 (7):\penalty0 991--1012,
  2011{\natexlab{b}}.

\bibitem[Chakravarty and Holden(1993)]{Chakravarty93}
S.~Chakravarty and C.~W. Holden.
\newblock An integrated model of market and limit orders.
\newblock \emph{Journal of Financial Intermediation}, 4:\penalty0 213--241,
  1993.

\bibitem[Cont and {de Larrard}(2013)]{ContLarrard}
R.~Cont and A.~{de Larrard}.
\newblock Price dynamics in a markovian limit order market.
\newblock \emph{SIAM Journal on Financial Mathematics}, 4:\penalty0 1--25,
  2013.

\bibitem[Cont et~al.(2010)Cont, Stoikov, and Talreja]{ContStoikov}
R.~Cont, S.~Stoikov, and R.~Talreja.
\newblock A stochastic model for order book dynamics.
\newblock \emph{Opererations Research}, 58:\penalty0 549–563, 2010.

\bibitem[Farmer et~al.(2005)Farmer, Patelli, and Zovko]{Farmer:2005predictive}
J.~D. Farmer, P.~Patelli, and I.~I. Zovko.
\newblock The predictive power of zero intelligence in financial markets.
\newblock \emph{Proceedings of the National Academy of Sciences of the United
  States of America}, 102\penalty0 (6):\penalty0 2254--2259, 2005.

\bibitem[Foucault et~al.(2005)Foucault, Kadan, and Kandel]{Foucault05}
T.~Foucault, O.~Kadan, and E.~Kandel.
\newblock Limit order book as a market for liquidity.
\newblock \emph{The Review of Financial Studies}, 18\penalty0 (4), 2005.

\bibitem[Gareche et~al.(2013{\natexlab{a}})Gareche, Disdier, Kockelkoren, and
  Bouchaud]{Gareche13}
A.~Gareche, G.~Disdier, J.~Kockelkoren, and J.-P. Bouchaud.
\newblock Fokker-planck description of the queue dynamics of large-tick stocks.
\newblock \emph{Phys. Rev. E}, 88:\penalty0 032809, 2013{\natexlab{a}}.

\bibitem[Gareche et~al.(2013{\natexlab{b}})Gareche, Disdier, Kockelkoren, and
  Bouchaud]{Gareche:2013fokker}
A.~Gareche, G.~Disdier, J.~Kockelkoren, and J.~P. Bouchaud.
\newblock Fokker--{P}lanck description for the queue dynamics of large tick
  stocks.
\newblock \emph{Physical Review E}, 88\penalty0 (3):\penalty0 032809,
  2013{\natexlab{b}}.

\bibitem[Gerig(2007)]{Gerig:2007theory}
A.~N. Gerig.
\newblock \emph{A {T}heory for {M}arket {I}mpact: {H}ow {O}rder {F}low
  {A}ffects {S}tock {P}rice}.
\newblock PhD thesis, University of Illinois at Urbana-Champaign, Champaign,
  IL, USA, 2007.

\bibitem[Glosten and Milgrom(1985)]{glosten}
L.~Glosten and P.~Milgrom.
\newblock Bid, ask and transaction prices in a specialist market with
  heterogeneously informed traders.
\newblock \emph{Journal of Financial Economics}, 14:\penalty0 71--100, 1985.

\bibitem[Gould and Bonart(2015)]{Gould:2015imbalance}
M.~D. Gould and J.~Bonart.
\newblock Queue imbalance as a one-tick-ahead price predictor in a limit order
  book.
\newblock \emph{arXiv:1512.03492}, 2015.

\bibitem[Gould et~al.(2013)Gould, Porter, Williams, McDonald, Fenn, and
  Howison]{Gould13}
M.~D. Gould, M.~A. Porter, S.~Williams, M.~McDonald, D.~J. Fenn, and S.~D.
  Howison.
\newblock Limit order books.
\newblock \emph{Quantitative Finance}, 13\penalty0 (11):\penalty0 1709--1742,
  2013.

\bibitem[Hautsch(2011)]{Hautsch:2011econometrics}
N.~Hautsch.
\newblock \emph{Econometrics of financial high-frequency data}.
\newblock Springer Science \& Business Media, 2011.

\bibitem[Ho and Stoll(1981)]{HO81}
T.~Ho and H.~R. Stoll.
\newblock Optimal dealer pricing under transactions and return uncertainty.
\newblock \emph{Journal of Financial Economics}, 9:\penalty0 47--73, 1981.

\bibitem[Huang and Polak(2011)]{Huang:2011LOBSTER}
R.~Huang and T.~Polak.
\newblock {LOBSTER: L}imit order book reconstruction system.
\newblock Technical report, Humboldt-Universit{\"a}t zu Berlin{,} available at
  \url{http://papers.ssrn.com/sol3/papers.cfm?abstract_id=1977207}, 2011.

\bibitem[Huang et~al.(2015)Huang, Lehalle, and Rosenbaum]{queue-reactive}
W.~Huang, C.-A. Lehalle, and M.~Rosenbaum.
\newblock Simulating and analyzing order book data: The queue-reactive model.
\newblock \emph{Journal of the Americal Statistical Association}, 110:\penalty0
  107--122, 2015.

\bibitem[Kirilenko et~al.(2014)Kirilenko, Kyle, Mehrdad, and
  Tugkan]{KirilenkoKyle}
A.~Kirilenko, A.~Kyle, S.~Mehrdad, and T.~Tugkan.
\newblock The flash crash: The impact of high frequency trading on an
  electronic market, 2014.

\bibitem[Kirilenko and Lamacie(2015)]{Kirilenko:2015latency}
A.~A. Kirilenko and G.~Lamacie.
\newblock Latency and asset prices.
\newblock \url{http://papers.ssrn.com/sol3/papers.cfm?abstract_id=2546567},
  2015.

\bibitem[Lee et~al.(2013)Lee, Eom, and Park]{Lee:2013microstructure}
E.~J. Lee, K.~S. Eom, and K.~S. Park.
\newblock Microstructure-based manipulation: Strategic behavior and performance
  of spoofing traders.
\newblock \emph{Journal of Financial Markets}, 16\penalty0 (2):\penalty0
  227--252, 2013.

\bibitem[Menkveld and Yueshen(2013)]{menkveld}
A.~Menkveld and B.~Yueshen.
\newblock Anatomy of the flash crash.
\newblock \url{http://papers.ssrn.com/sol3/papers.cfm?abstract_id=2243520},
  2013.

\bibitem[Mike and Farmer(2008)]{Mike:2008empirical}
S.~Mike and J.~D. Farmer.
\newblock An empirical behavioral model of liquidity and volatility.
\newblock \emph{Journal of Economic Dynamics and Control}, 32\penalty0
  (1):\penalty0 200--234, 2008.
\newblock ISSN 0165-1889.

\bibitem[Ohara and Oldfield(1986)]{Ohara86}
M.~Ohara and G.~Oldfield.
\newblock Microeconomics of market making.
\newblock \emph{Journal of Financial and Quantitative Analysis}, 21:\penalty0
  361--367, 1986.

\bibitem[{Ro\c su}(2014)]{Rosu14}
I.~{Ro\c su}.
\newblock Liquidity and information in order driven markets.
\newblock \url{http://papers.ssrn.com/sol3/papers.cfm?abstract_id=1286193},
  2014.

\bibitem[Ro{\c s}u(2009)]{Rosu09}
I.~Ro{\c s}u.
\newblock A dynamical model of the limit order book.
\newblock \emph{The Review of Financial Studies}, 22\penalty0 (11):\penalty0
  4601--4641, 2009.

\bibitem[Smith et~al.(2003)Smith, Farmer, Gillemot, and
  Krishnamurthy]{Smith:2003statistical}
E.~Smith, J.~D. Farmer, L.~Gillemot, and S.~Krishnamurthy.
\newblock Statistical theory of the continuous double auction.
\newblock \emph{Quantitative Finance}, 3\penalty0 (6):\penalty0 481--514, 2003.

\bibitem[Weber and Rosenow(2005)]{Rosenau}
P.~Weber and B.~Rosenow.
\newblock Order book approach to price impact.
\newblock \emph{Quantitative Finance}, 5:\penalty0 357--364, 2005.

\end{thebibliography}

\end{document}